\documentclass[a4paper,11pt]{article}
\usepackage{jheppub} 

\usepackage[x11names,dvipsnames,table]{xcolor}
\usepackage{braket}
\usepackage{tikz}
\usetikzlibrary {arrows.meta}
\usepackage{dynkin-diagrams}

\usepackage{booktabs}
\usepackage{adjustbox}

\definecolor{mygrey}{RGB}{220, 220, 220}

\title{\boldmath On massless RSOS flows}

\author[a]{Máté Lencsés}
\author[a,b]{Karolina S. Trombitás}
\affiliation[a]{HUN-REN Wigner Research Centre for Physics,\\Konkoly-Thege Mikl\'os u. 29-33, 1121 Budapest, Hungary}
\affiliation[b]{Department of Theoretical Physics, Institute of Physics,\\Budapest University of Technology and Economics,\\Műegyetem rkp. 3., H-1111 Budapest, Hungary}

\abstract{We reconsider deformations of massless RSOS scattering theories, leading to various RG flows. We detail the ultraviolet and infrared limits both analytically and numerically. For the flows starting from the parafermionic Liouville theory we show that the reflection amplitude quantisation works also for shifted kernels. Different deformations lead to different UV completions and the UV perturbation determines the residual category of commuting topological lines along the flows. This in turn constrains the nature of irrelevant perturbations in the infrared: scalar CDD deformations correspond to generalised $T\bar T$ perturbations, while the RSOS-type matrix structure leads to the presence of $\Phi_{3,1}$. The scalar CDD deformation is expected to leave the symmetry category intact: we show this for the pure scalar deformation and for the $n=2$ case of the parafermionic flows and argue that it holds in general.}

\begin{document}
\maketitle
\flushbottom

\section{Introduction}
\label{sec:intro}

Two-dimensional renormalisation group (RG) flows play a crucial role in our understanding of quantum field theories at various energy scales. Under standard assumptions, scale-invariant fixed points are characterised by conformal field theories (CFT)~\cite{Polchinski:1987dy,NAKAYAMA20151}, which can be solved in many two-dimensional cases due to the infinite number of conserved quantities~\cite{Belavin:1984vu}. The RG flows can be thought of as relevant perturbations of such CFTs as the ultraviolet (UV) fixed point.

The relevant perturbations can be either massive or massless. In the massive case, the infrared (IR) fixed point is trivial in the sense that the correlation length vanishes exponentially as we near the fixed point. On the other hand, massless flows flow to other interacting CFTs.

Of particular interest are the so-called integrable flows, where one can find infinitely many conserved charges along the flow. Such theories can be solved by means of the integrable bootstrap program: elementary excitations, their scattering matrix and operator matrix elements can be obtained analytically~\cite{1979AnPhy.120..253Z,Smirnov:1992vz}.

Moreover, using the scattering data and space-time duality in Euclidean formulation, one can treat these models in finite volume, which acts as an energy scale: small volume is related to the UV while large volume to the IR limit. The method is called the thermodynamic Bethe ansatz (TBA)~\cite{Zamolodchikov:1989cf}, which was originally developed for massive scattering and RG flows.

However, various massless scattering theories and TBA equations have been studied~\cite{Zamolodchikov:1991vx,Zamolodchikov:1991vg,Fateev:1991bv,1992NuPhB.379..602Z,1993IJMPA...8.5751F}, some of which are conjectured TBA equations and some are derived directly from scattering of left- and right-moving particles.
In the fixed points, the scattering between left- and right-moving particles is trivial. Given the context of generalised $T\bar T$ deformations~\cite{2004hep.th....1146Z,2016JHEP...10..112C,SmirnovZamolodchikov2017NuPhB.915..363S}, it is natural to introduce a non-trivial scattering between the left- and right-moving particle, however this procedure goes back to~\cite{1992NuPhB.379..602Z,1993IJMPA...8.5751F}. Such a deformation of the scattering leads to new RG flows, that enter to the same fixed point in the IR, i.e. the deformations lead to irrelevant perturbations of the fixed point. This change in perspective opened a way to the study of new massless RG flows~\cite{2022JHEP...08..179A,AhnBajnok2024,2025JHEP...12..103A,2026arXiv260628783A,2026arXiv260824385A}: using the deformed TBA equations, one can analyse such flows exactly. In many cases the deformed TBA can be solved deep in the UV and allows to find the UV limit, while in certain cases Hagedorn-like singularities occur, preventing to reach the UV limit~\cite{2021JHEP...10..062C,2022JHEP...08..179A,2024JHEP...06..157A}.

One such program was recently carried out in~\cite{AhnBajnok2024} for the so-called restricted solid-on-solid (RSOS) scattering theories~\cite{Bernard:1990cw}, allowing for the derivation of previously conjectured massless TBA systems~\cite{Zamolodchikov:1991vx,Zamolodchikov:1991vg,Fateev:1991bv} and predicting new UV finite RG flows.

Another recent advancement of the field has been sparked by the seminal paper~\cite{2015JHEP...02..172G} extending the notion of ordinary group-like symmetry of quantum theory to more general algebraic structures such as higher-form, non-invertible or higher-categorical symmetries, see~\cite{2023arXiv230800747S,2024PhR..1063....1S} for recent reviews. In particular, it led to the understanding of the role of non-invertible symmetries in 1+1 dimensions see e.g.~\cite{2019JHEP...01..026C,2020arXiv200808598A,2024PhRvL.133r1601C,2025JHEP...03..204C}.

In the fixed-point CFTs such symmetry is generated by topological lines. In the simplest cases, the topological lines are exactly the Verlinde lines, in one-to-one correspondence with the available representations of the fixed-point extended symmetry~\cite{PetkovaZuber2001PhLB..504..157P}. It is natural to expect that RG flows carry some subset of the symmetry category of their fixed points~\cite{TanakaNakayama2024JHEP...11..137N,2025PhRvL.135b1602A,2026arXiv260309591B}. In turn, this restricts the possible irrelevant perturbations of the IR fixed point given the UV perturbation and vice versa.

In this work, we revisit the RG flows generated by various deformations of the massless RSOS scattering. We detail the IR and UV limits both numerically and analytically. In particular, we demonstrate the Liouville reflection amplitude quantisation for the flows with parafermions. Combining the residual non-invertible symmetries determined using the ideas of ~\cite{TanakaNakayama2024JHEP...11..137N,2026arXiv260309591B} with our infrared TBA analysis, we show that the structure of the S-matrix deformation is mirrored in the irrelevant operators at the IR fixed point. Using the IR behaviour of the effective central charge, we further verify that pure CDD deformations indeed change the coupling of the irrelevant $T\bar T$ deformation.

The structure of the paper is as follows. In Section~\ref{sec:TBAs} we summarise the deformations of the RSOS TBA systems based on~\cite{AhnBajnok2024}, then we detail the UV limit in Section~\ref{sec:UV} and the IR in Section~\ref{sec:IR} focusing on the context of non-invertible symmetries. Finally, we conclude in Section~\ref{sec:outro}. Detailed numerical results in the UV and IR limits are relegated to the Appendices.

\section{Massless flows, RSOS and deformations}\label{sec:TBAs}

\subsection{RSOS scattering}

The restricted solid-on-solid (RSOS) scattering~\cite{Bernard:1990cw} is based on a field theory with multiple ground states, whose vacua can be represented by the $A$-type Dynkin diagram. In the $\text{RSOS}_p$ theory, there are $p$ degenerate ground states, labeled by Latin letters $a=0,\frac{1}{2},\dots,\frac{p+1}{2}-1$. Excitations are kinks interpolating between ground states $a$ and $a\pm\frac{1}{2}$. We summarise the RSOS scattering and the TBA equations using the notation of~\cite{AhnBajnok2024}. 

A kink excitation with rapidity $\theta$ can be denoted pictorially with:
\begin{equation}
    a\, \bigg\uparrow\, b = K_{ab}(\theta),\qquad a,b=0,\frac{1}{2},\dots,\frac{p+1}{2}-1\;,\quad \text{s.t.}\quad |a-b|=\frac{1}{2}\,.
\end{equation}
The kinks can scatter on each other, with a non-diagonal scattering matrix that respects the RSOS condition and the Yang--Baxter equation. Let us denote an $S$-matrix element $S_p(\theta)_{dc}^{ab}$ as the amplitude of the following two-kink scattering process:
\begin{equation}
    K_{dc}(\theta_1)+K_{cb}(\theta_2) \rightarrow K_{da}(\theta_2) + K_{ab}(\theta_1)\,.
\end{equation}
The $S$-matrix element takes the form:
\begin{equation}\label{eq:Skink}
    S_{p}(\theta)^{ab}_{dc} = U(\theta) (X^{ac}_{db})^{\frac{i\theta}{2\pi}}\left[(X^{ac}_{db})^{\frac{1}{2}}\sinh\left(\frac{\theta}{p+1}\right)\delta_{db}+\sinh\left(\frac{i\pi-\theta}{p+1}\right)\delta_{ac}\right]\,,
\end{equation}
where $U(\theta)$ is the scalar factor:
\begin{equation}
    U(\theta) = \frac{1}{\sinh\left(\frac{\theta-i\pi}{p+1}\right)}\exp\left[\int_{-\infty}^{\infty}\frac{dk}{k} \frac{\sinh\frac{k\pi p}{2}}{2\sinh\frac{k\pi(p+1)}{2}\cosh\frac{k\pi}{2}}e^{ik\theta}\right]\,,
\end{equation}
and
\begin{equation}
    X_{db}^{ac} = \frac{[2a+1][2c+1]}{[2d+1][2b+1]}\,,\qquad [x]=\frac{q^x-q^{-x}}{q-q^{-1}}\,,\quad \text{with}\quad q=-\exp\left(\frac{-i\pi}{p+1}\right)\,.
\end{equation}

In the ultraviolet limit, this theory is the perturbation of the unitary minimal model CFT $\mathcal{M}_{p+1}$, with central charge $c=1-\frac{6}{(p+1)(p+2)}$, the perturbing field is the relevant $\Phi_{1,3}$ of left and right conformal weights $h=\bar h=\frac{p}{p+2}$. The action of the theory reads as
\begin{equation}
    \mathcal{S}_{\text{MA}_p^-} = \mathcal{S}_{\mathcal{M}_{p+1}} + \lambda\int d^2x \Phi_{1,3}\,,
\end{equation}
with $\lambda<0$. Changing the sign of the coupling leads to a massless flow to $\mathcal{M}_p$ in the infrared~\cite{Zamolodchikov:1991vx}. 

The UV CFT can be viewed as the $\lambda\rightarrow 0^-$ limit. In this limit the scattering states become right- and left-moving massless particles, with dispersion relation $E=\pm P$ for right- and left-movers respectively (R and L for short). The states can be thought of as ultra-relativistic particles with rescaled rapidities $\theta_i \rightarrow \hat \theta_{i} \pm \Lambda$ for R/L respectively, with $\Lambda \rightarrow \infty$ and $M\rightarrow0$, while $Me^{\Lambda}$ fixed. The dispersion relation then reads as:
\begin{equation}\label{eq:masslessdisp}
    E=\hat{M} e^{\pm\hat\theta},\qquad + \text{for R and} - \text{for L}\,.
\end{equation}

In this limit, the right-right and left-left scattering amplitudes remain the same as the massive case:
\begin{equation}
    S_p^{RR}(\theta)=S_p^{LL}(\theta)=S_p(\theta)\,
\end{equation}
while the left-right scattering is independent of the rapidity difference:
\begin{equation}
    S_p^{LR} = \lim_{\theta\rightarrow\infty} S_p(\theta)\,,
\end{equation}
leading to a constant phase shift. The trivial left-right scattering is nothing else, but the manifestation of the chiral factorisation in the conformal field theory.

\subsection{RSOS TBA}

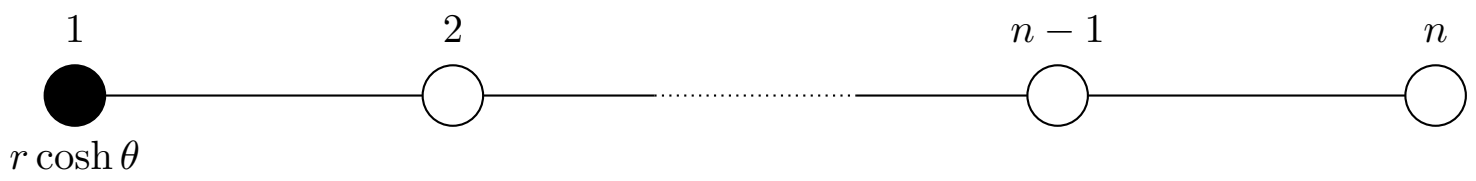
\begin{figure}[t]
    \centering
    \pgfkeys{/Dynkin diagram,edge length=2.5cm,fold radius=.5cm}
\begin{tikzpicture}
\dynkin[labels*={1,2,n-1,n},labels={r \cosh \theta},root radius=0.2cm]A{*o.oo}
\end{tikzpicture}
    \caption{Adjacency diagram for the TBA corresponding to the massive $\text{RSOS}_{n+1}$ scattering theory.}
    \label{fig:massiveRSOS}
\end{figure}

The massive version of the RSOS$_{n+1}$ TBA system was conjectured by Zamolodchikov in~\cite{Zamolodchikov:1991vh} to be\footnote{Here $n$ TBA nodes correspond to $p=n+1$ RSOS vacua, so the UV limit is $\mathcal{M}_{n+2}$. Later we choose the number of TBA nodes in a specific way: the organising principle is the IR minimal model which is chosen to be $\mathcal M_{n+1}$ for a related set of models.}
\begin{equation}
    \epsilon_a(\theta) = \delta_{a1}M R \cosh\theta-\sum_{b=1}^n \mathbb{I}_{ab}\varphi\star \log\left( 1+e^{-\epsilon_b}\right)(\theta)\,,\quad a=1,2,\dots,n\,,
\end{equation}
where $\epsilon _a$ are the pseudo energies and $\varphi$ is the "universal" kernel:
\begin{equation}
    \varphi(\theta) = \frac{1}{\cosh \theta}\,,
\end{equation}
and $\star$ denotes the convolution $f\star g(\theta)=\frac{1}{2\pi}\int_{-\infty}^{\infty}d\theta'f(\theta-\theta')g(\theta')$ while $\mathbb I$ is the adjacency matrix of the TBA system. In the case of the massive RSOS it is given in Figure~\ref{fig:massiveRSOS}.
The effective central charge function can be computed from the pseudo-energy:
\begin{equation}
    c(r) = \frac{3r}{\pi^2} \int_{-\infty}^\infty d\theta \cosh(\theta) \log(1+e^{-\epsilon_1})(\theta)\,,
\end{equation}
where $r=MR$ is the dimensionless volume.

\begin{figure}
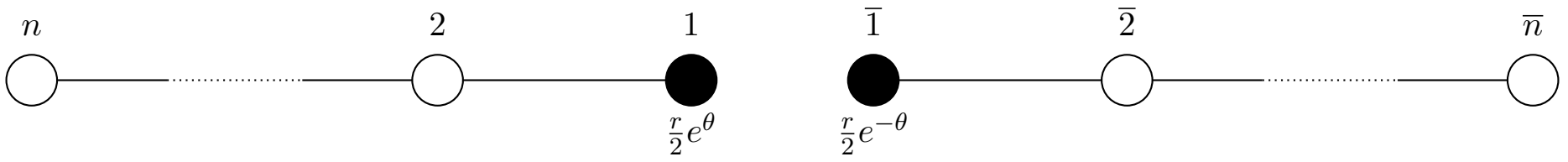

    \centering
    \pgfkeys{/Dynkin diagram,edge length=2cm,fold radius=.5cm}

\dynkin[labels*={n,2,1},labels={,,\large{\frac{r}{2}e^{\theta}}},root radius=0.2cm]A{o.o*}\qquad
\dynkin[labels*={\overline{1},\overline{2},\overline{n}},labels={\frac{r}{2}e^{-\theta},,},root radius=0.2cm]A{*o.o}
    \caption{In the massless limit the TBA equations for the right- and left movers decouple, leading to two identical TBAs, apart from the source terms.}
    \label{fig:masslessRSOS}
\end{figure}

In the massless case the left-right scattering is independent of the rapidity difference. This means that when writing the TBA equations for right and left movers, the left-right kernel vanishes. Therefore, the TBA decouples to two independent systems for right and left movers. These systems are equivalent to the massive RSOS TBA, except the source terms are changed to $\hat M R e^{\pm\theta}$. The corresponding TBA equation can be summarised as:
\begin{equation}
\begin{aligned}
    \epsilon^{+}_a(\theta) & = \delta_{a1}\frac{\hat{M}R}{2} e^{+\theta}-\sum_{b=1}^n \mathbb{I}_{ab}\varphi\star \log\left( 1+e^{-\epsilon^{+}_b}\right)(\theta)\,,\quad a=1,2,\dots,n\,,\\
    \epsilon^{-}_a(\theta) & = \delta_{ a \bar 1}\frac{\hat{M}R}{2} e^{-\theta}-\sum_{b=\bar 1}^{\bar n} \mathbb{I}_{ab}\varphi\star \log\left( 1+e^{-\epsilon^{-}_b}\right)(\theta)\,,\quad a=\bar 1,\bar 2,\dots,\bar n\,,
    \end{aligned}
\end{equation}
where now the adjacency matrix can be read off from Figure~\ref{fig:masslessRSOS}.

The effective central charge is given in terms of the source nodes:
\begin{equation}
    c(r) = \frac{3r}{2\pi^2} \int_{-\infty}^\infty d\theta \left[e^\theta \log\left(1+e^{-\epsilon_1(\theta)}\right)+e^{-\theta} \log\left(1+e^{-\epsilon_{\bar 1}(\theta)}\right)\right]\,,
\end{equation}
where $r=\hat MR$ is the dimensionless volume.

\subsection{Irrelevant deformations of the massless scattering}

In this work we reconsider the irrelevant deformations of the massless RSOS scattering recently studied in~\cite{AhnBajnok2024} that lead to new RG flows. We complete the picture initiated in~\cite{AhnBajnok2024}. In particular, we are interested in RG flows that have $\mathcal{M}_{n+1}$ as \emph{infrared} fixed points generated by deformations to the scattering between right- and left-moving particles. These naturally fit into the generalised $T \bar T$ picture~\cite{SmirnovZamolodchikov2017NuPhB.915..363S,2021JHEP...10..062C}.

The generalised $T \bar T$ deformations lead to CDD deformations of the $S$-matrix~\cite{2012JHEP...09..133D,2013arXiv1305.1278C,SmirnovZamolodchikov2017NuPhB.915..363S}. Such deformations usually lead to UV-incomplete theories. However, careful tuning of the couplings of higher charges might lead to UV-complete theories~\cite{2019JHEP...11..120C,2020PhRvL.124t0601H,2021JSMTE2021k3104L}.

In the massless scattering problem, one can also deform the trivial scattering between right- and left-movers to a non-diagonal scattering, as a solution to the Yang-Baxter equations. This was initiated in~\cite{1993IJMPA...8.5751F}, where a shifted $S$-matrix was conjectured between right- and left-movers.

Along these lines, one can introduce two types of deformations~\cite{AhnBajnok2024}:
i) identity type with CDD factors ii) non-diagonal RSOS type, with or without CDD. In the second scenario, the RL and LR $S$-matrices might have oppositely shifted rapidity arguments. Below we list the resulting TBA systems of such deformations~\cite{1993IJMPA...8.5751F,AhnBajnok2024}.

\begin{itemize}
    \item Identity+CDD deformation leads to the TBA system in Figure~\ref{fig:iddeformed}, that we call double $A_{n-1}$, which was originally conjectured in~\cite{Zamolodchikov:1991vg} and denoted by $\mathcal{M}A^+(n-1,1)$.
    \item Same $S$-matrix deformation leads to the $D_n$ TBA system in Figure~\ref{fig:Dndiag}, see~\cite{AhnBajnok2024}. This TBA was conjectured in~\cite{Fateev:1991bv} and was denoted there as $H^{(\pi)}_n$.
    \item Same $S$-matrix with additional CDD factor leads to the $\hat D_n$ diagram, see eq.~\eqref{eq:Dhat}\footnote{We note that in this case the hat does not refer to the affine $D$ Dynkin diagram. It symbolizes the additional kernel between the source nodes.}. This flow turns out to be UV incomplete, but it has interesting features in the IR.
    \item Shifted $S$-matrix leads to Zamolodchikov's flows connecting unitary minimal models~\cite{1993IJMPA...8.5751F}, usually denoted as $\mathcal MA_{n+1}^+$.
    \item Shifted $S$-matrix + CDD deformation leads to the affine $\hat A _n$ TBA system~\cite{AhnBajnok2024}.
\end{itemize}

Adding more CDD factors leads to theories that are not UV complete and lead to Hagedorn-like singularities, which we verified numerically.

These deformations can also be considered as irrelevant perturbations of the infrared theory:
\begin{equation}
    \mathcal S_{IR} = \mathcal S_{\mathcal M_{n+1}} + \sum_{i\,\text{irrelevant}} \lambda_i \int d^2 x \Phi_i + \sum_{s\geq1}\beta_s\int d^2 x [T\bar T]_s\,,
\end{equation}
where we separated the family that are known as generalised $T\bar T$ deformations that generate additional CDD factors to the $S$-matrix~\cite{SmirnovZamolodchikov2017NuPhB.915..363S}.

To be more concrete, the right-left CDD factor becomes:
\begin{equation}\label{eq:CDDexp}
    S^{(\beta)\text{RL}}_{\text{CDD}}(\theta)=S^{(\beta)\text{LR}}_{\text{CDD}}(-\theta) =\prod_{s=1}^{\infty} \exp [i\hat g_s e^{s\theta}]\,,
\end{equation}
where $\beta$ stands for the collection $\beta_s$ and $\hat g_s = -\beta_s\hat M^{2s}$ with $\hat M$ being a mass scale defined in the massless-scattering limit in~\eqref{eq:masslessdisp}. The CDD factor in~\cite{AhnBajnok2024} is
\begin{equation}\label{eq:CDDfactor}
    S_{\text{CDD}}^{RL}(\theta)=S_{\text{CDD}}^{LR}(-\theta) = -\tanh \left( \frac{\theta+\alpha}{2}-\frac{i \pi}{4}\right)\,,
\end{equation}
where $\alpha$ is an arbitrary real parameter, leading to odd terms only in~\eqref{eq:CDDexp}.\footnote{Note that our convention differs from~\cite{AhnBajnok2024} in the sign of $\alpha$.}

It is also clear that pure generalised $T\bar T$ cannot generate a non-diagonal RSOS scattering at least one non-trivial primary should be added. The perturbation was found to be $\Phi_{3,1}$ for the flows between minimal models in~\cite{Zamolodchikov:1987,Ludwig:1987gs} and in~\cite{Fateev:1991bv} for the $D_n$ type TBA, see also~\cite{Feveratietal1996PhLB..374...64F} for a more detailed analysis. Moreover, it is also possible to fix the potential IR perturbations from the UV point of view using non-invertible symmetries: once the relevant UV perturbation is identified, the non-invertible symmetry along the flow can be determined and has to be compatible with the IR perturbation~\cite{TanakaNakayama2024JHEP...11..137N,2025PhRvL.135b1602A,2026arXiv260309591B}. We detail on this approach in Section~\ref{sec:IR}.

In the subsequent sections we analyze the various deformations in detail. We recall the results for the UV fixed point CFT and the relevant perturbation that initiates the RG flow, and support it with extensive numerics. The IR fixed point is always the corresponding unitary minimal model, although it is clear that they enter from different irrelevant deformations. This direction has to respect the non-invertible symmetry of the flows, that are in turn fixed by the topological lines commuting with the relevant perturbation in the UV. For completeness, we summarise the above properties of the various deformations in Table~\ref{tab:models}.

\begin{table}[ht]
    \centering
    \begin{tabular}{cccccc}
    \hline
        Deformation & Id.+CDD & $S$ & $S$+CDD &sh.$S$ & sh.$S$+CDD \\[-1em] \\ 
    \hline \\[-1em]
         Model & double $A_{n-1}$ & $D_n$ & $\hat D_n$ & massless $A_n$  & $\hat A_n$  \\[-1em] \\
    \hline diagram & \dynkin A{o.o**o.o} & \dynkin D{o.oo**} & \dynkin [fold]D{o.oo**} &\dynkin A{*o.o*}& \begin{DynkinDiagrams}{A{*o.o*}}\draw (0,0) arc (120:60:1.257);\end{DynkinDiagrams}    \\
    \hline \hline\\[-1em]
         UV & $\frac{su(2)_{n}\times su(2)_{n-1}}{su(2)_{2n-1}}$ & $\frac{su(2)_n}{u(1)_n}$ &\footnotesize no UV& $\mathcal{M}_{n+2}$  & PL${}_n$ \\[-1em] \\
   
     $c^{\text{eff}}_{\text{UV}}$ &$\frac{3(n-1)n(2n+3)}{(n+1)(n+2)(2n+1)}$ & $\frac{2n-2}{n+2}$ & \footnotesize no UV& $1-\frac{6}{(n+2)(n+3)}$  & $\frac{3n}{n+2}$ \\[-1em]\\
    & \checkmark & \checkmark & \checkmark & \checkmark & \checkmark 
     \\[-1em]\\
     \hline\\[-1em]
     
     UV pert.  & \footnotesize $\Phi^{0,0}_2$ & \footnotesize $V^{(\pi)}_{(0,\pm2)}$ & \footnotesize no UV& \footnotesize $\Phi_{1,3}$ &  \footnotesize $\eta\,\psi_1^\dagger \bar \psi_1^\dagger e^{-2b\phi}$ 
     \\[-1em] \\

     & \checkmark & \checkmark & \checkmark & \checkmark & ? 
     \\[-1em]\\
    \hline\\[-1em]
         IR & $\mathcal{M}_{n+1}$ & $\mathcal{M}_{n+1}$ & $\mathcal{M}_{n+1}$ & $\mathcal{M}_{n+1}$ & $\mathcal{M}_{n+1}$ \\[-1em] \\
         & \checkmark & \checkmark & \checkmark & \checkmark & \checkmark 
     \\[-1em]\\
    \hline\\[-1em]
        IR pert. & \footnotesize $T\bar{T}$ & \footnotesize $-\Phi_{3,1}$  & \footnotesize $-\Phi_{3,1}+ T\bar T$  & \footnotesize $\Phi_{3,1}+\sigma_n T\bar T$ & \footnotesize $\Phi_{3,1}+(\sigma_n+1) T\bar T$ \\[-1em]\\
        & \checkmark & ? & ? & ? & ? 
     \\[-1em]\\
    \hline\\[-1em]
    residual NIS & $\mathcal L^{(n+1)}_{(r,s)}$ & $\mathcal L^{(n+1)}_{(1,s)}$ & no UV & $\mathcal L^{(n+1)}_{(1,s)}$ & $\mathcal L^{(n+1)}_{(1,s)}$\\
    & \checkmark & \checkmark & \checkmark & \checkmark & ? 
     \\[-1em]\\
    \hline
    \end{tabular}
    \caption{Properties of the massless flows studied in this paper. $\mathcal L^{(n+1)}_{(r,s)}$ denotes the fusion category of the diagonal minimal model $\mathcal M_{n+1}$, while $\mathcal L^{(n+1)}_{(1,s)}$ is the subcategory commuting with $\Phi_{3,1}$. We emphasised with a question mark wherever the result needs further analysis: the correct massless ray and/or further topological lines in the $\hat A_n$ case (cf. Section~\ref{sec:hatAIR}), the status of the presence of the generalised $T\bar T$ family in the $D_n$ case (cf. Section~\ref{sec:dnIR}), and the sign pattern $\sigma_n=-\cos \left(\frac{n \pi}{2}\right)$ of the $T\bar T$ coupling in the massless $A_n$ (cf. Sections~\ref{sec:masslessAn} and ~\ref{sec:hatAIR}).
    }
    \label{tab:models}
\end{table}

\section{Ultraviolet limits}\label{sec:UV}

In this section, we present a detailed numerical account for the UV limit of TBA systems listed at the end of the previous section, with the exception of the well-established massless $A_n$ between minimal models.

\subsection{Double $A_{n-1}$: identity with CDD}

\begin{figure}
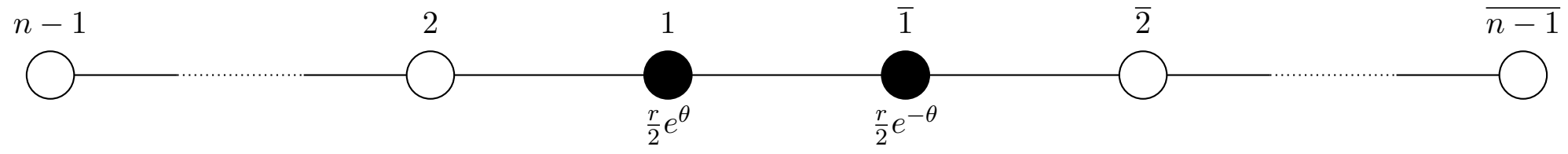

    \centering
        \pgfkeys{/Dynkin diagram,edge length=2cm,fold radius=.5cm}

\dynkin[labels*={n-1,2,1,\overline{1},\overline{2},\overline{n-1}},labels={,,\frac{r}{2}e^{\theta},\frac{r}{2}e^{-\theta},,},root radius=0.2cm]A{o.o**o.o}
    \caption{Adjacency diagram from identity deformed $S$-matrix deformation.}
    \label{fig:iddeformed}
\end{figure}

Introducing diagonal scattering with a universal CDD factor~\eqref{eq:CDDfactor} with $\alpha=0$ between the left- and right-movers leads to a TBA which is governed by two coupled massless $A_{n-1}$ diagrams as shown in Figure~\ref{fig:iddeformed}, and it reads as:
\begin{equation}
    \begin{aligned}
        \epsilon_1 + \varphi \star L_2 + \varphi \star L_{\overline{1}} &= \frac{r}{2} e^\theta \\
        \epsilon_{\overline{1}} + \varphi \star L_1 + \varphi \star L_{\overline{2}} &= \frac{r}{2} e^{-\theta} \\
        \epsilon_a + \varphi \star L_{a-1} + \varphi \star L_{a+1} &= 0 \hspace{1 cm} \text{for}\, a = 2,...,n-2, \overline{2}, ...,\overline{n-2} \\
        \epsilon_{n-1} + \varphi \star L_{n-2} &= 0 \\
        \epsilon_{\overline{n-1}} + \varphi \star L_{\overline{n-2}} &= 0\,,
    \end{aligned}
    \label{eq:identityTBA}
\end{equation} 
where we introduced the notation $L_i=\log(1+e^{-\epsilon_i})$.

This TBA system, along with its more general form, was conjectured in~\cite{Zamolodchikov:1991vg}, where the UV expansion of the effective central charge was determined
\begin{equation}\label{eq:IdUV}
    \tilde{c}(r) =\frac{3(n-1)n(2n+3)}{(n+1)(n+2)(2n+1)} + \mathcal{B} r^2 +  \sum_{k=2}^{\infty} \tilde{c}_k r^{4k/(2n+1)}\,,
\end{equation}
where
\begin{equation}
\mathcal{B}=
    \begin{cases}
        0 & n \,\text{odd}\\
        \frac{3}{2\pi} & n \,\text{even}\;.
    \end{cases}
\end{equation}

\begin{figure}[t]
    \centering
    \includegraphics[width=0.7\linewidth]{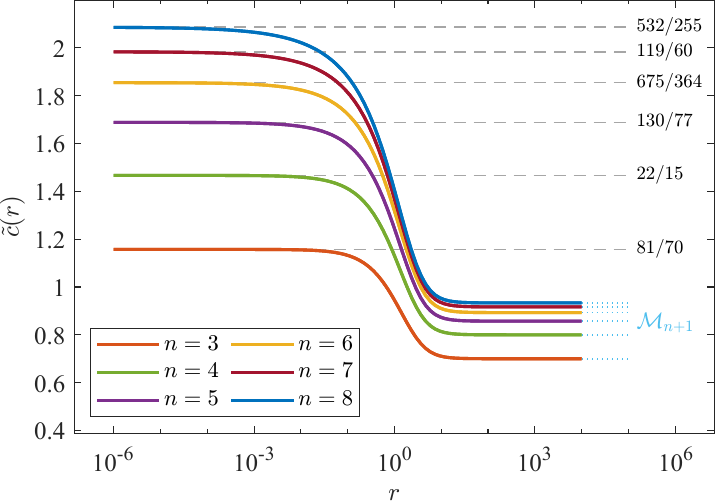}
    \caption{Effective central charge from the double $A_{n-1}$ TBA originating from the identity deformed $S$-matrix.}
    \label{fig:IdentityPlot}
\end{figure}

We performed extensive numerical computations for various $n$s by iterating the TBA in the UV limit to verify these results. Figure~\ref{fig:IdentityPlot} summarises the dimensionless volume dependence of the effective central charges. It is clear that the UV central charge is predicted by~\eqref{eq:IdUV}. The value of UV central charge corresponds to the coset $\frac{su(2)_{n}\times su(2)_{n-1}}{su(2)_{2n-1}}$, while the perturbation generating the flow has conformal dimensions $h=\bar h = \frac{2n-1}{2n+1}$~\cite{Zamolodchikov:1991vg}, which is in excellent agreement with our numerical data that we present in Table~\ref{tab:identity_fitresult} in Appendix~\ref{app:A} for $n=3,4$, where the fitted exponent $A$ tests $2-2h$.

\subsection{$D_n$: same $S$-matrix}

\begin{figure}
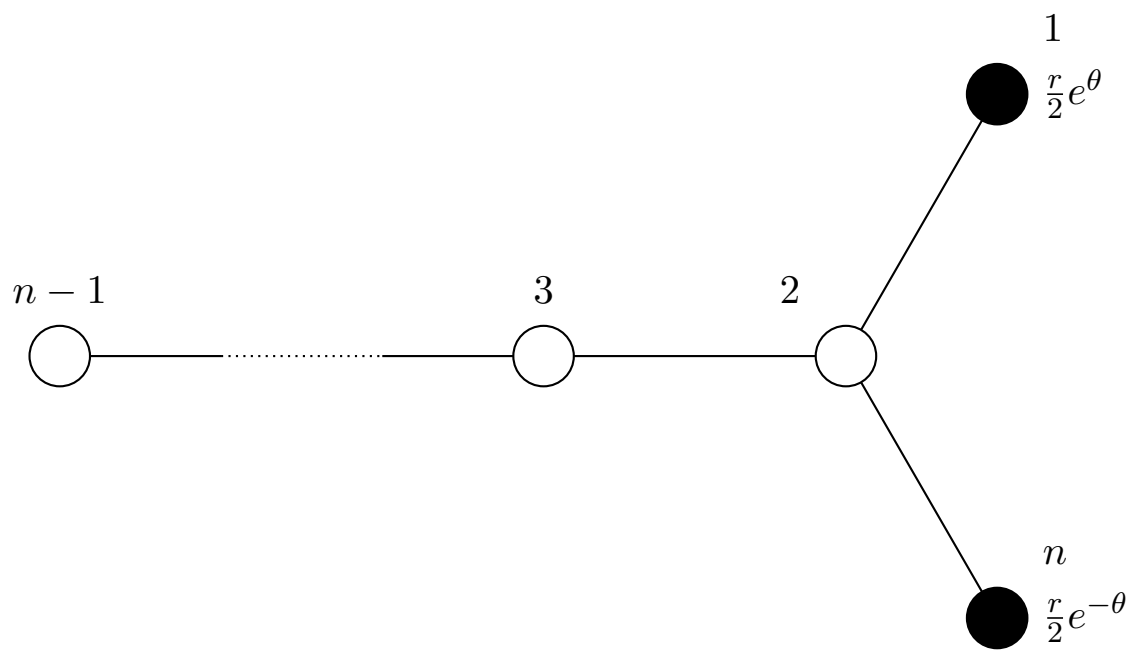

    \centering
            \pgfkeys{/Dynkin diagram,edge length=2cm,fold radius=.5cm}

\dynkin[labels*={n-1,3,2,1,n},labels={,,,\frac{r}{2}e^{\theta},\frac{r}{2}e^{-\theta}},root radius=0.2cm]D{o.oo**}
    \caption{$D_n$ adjacency diagram for the same $S$-matrix deformation.}
    \label{fig:Dndiag}
\end{figure}

This TBA system reads as:

\begin{equation}
    \begin{aligned}
        \epsilon_1(\theta) + (\varphi \star L_2)(\theta) &= \frac{r}{2}e^\theta \\
        \epsilon_n(\theta) + (\varphi \star L_2)(\theta) &= \frac{r}{2}e^{-\theta} \\
        \epsilon_2(\theta) + (\varphi \star L_1)(\theta) + (\varphi \star L_3)(\theta) + (\varphi \star L_n)(\theta) &= 0 \\
        \epsilon_a(\theta) + (\varphi \star L_{a-1})(\theta) + (\varphi \star L_{a+1})(\theta) &= 0\,, \hspace{1cm}\text{for }\, a = 3,...,n-2 \\ 
        \epsilon_{n-1}(\theta) + (\varphi \star L_{n-2})(\theta) &= 0\,,
    \end{aligned}
    \label{eq:sameSTBA}
\end{equation}
encoded in the $D_n$ diagram presented in Figure~\ref{fig:Dndiag}.

This TBA system was first conjectured in~\cite{Fateev:1991bv} and shown to be generated by the "same $S$-matrix" deformation of the massless RSOS scattering theory in~\cite{AhnBajnok2024}.

Based on the periodicity property of the corresponding $Y$-system, the UV expansion of the effective central charge of this system has been conjectured in~\cite{Fateev:1991bv}, and has the form:

\begin{equation}
    \tilde{c}(r) = \frac{2n-2}{n+2} - c^{(n)}_{\log} r^2 \log(r) + \sum_{k=2}^\infty \tilde{c}^{(n)}_k r^{2k/n}
    \label{eq:sameSc_UVexpansion}
\end{equation}
where the coefficient of the logarithmic term has been computed:
\begin{equation}
    c^{(n)}_{\log}=
    \begin{cases}
          \frac{3(n-2)}{2 \pi^2 n} &  n \, \text{even} \\
         \frac{3(n-1)}{2 \pi^2 n} &  n \, \text{odd}\,.
    \end{cases}
\end{equation}

\begin{figure}
    \centering
    \includegraphics[width=0.7\linewidth]{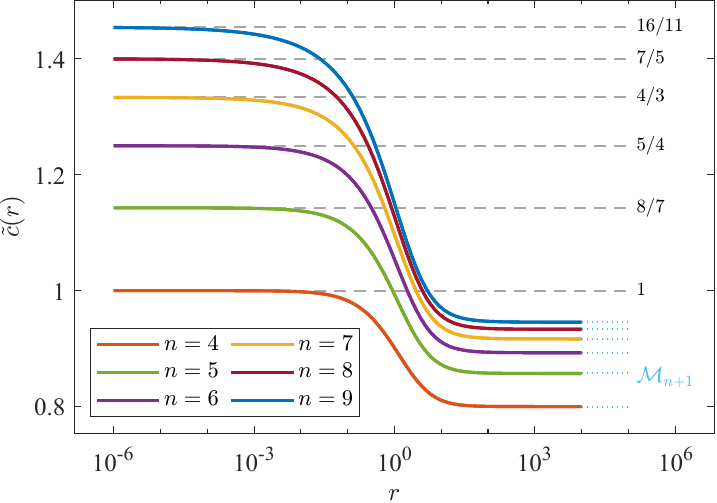}
    \caption{Effective central charge for the same $S$-matrix deformation, i.e. the massless $D_n$ Dynkin diagram TBA.}
    \label{fig:sameSplot}
\end{figure}

To our knowledge, no numerical analysis has been carried out to test the UV expansion. We numerically solved the above TBA system, for different values of $n$, the resulting effective central charge functions are presented in Figure~\ref{fig:sameSplot}. For the first two nontrivial cases ($n=4,5$) we also carried out numerical fits using the following functional form:
\begin{equation}
    \tilde{c}(r) = c_n^{\rm UV} + B r^2 \log(r) + \sum_{k=1}^{k_{\rm max}} \tilde{c}_k \left(r^{A}\right)^k\,,
    \label{eq:sameS_fitfunc}
\end{equation}
where the fitted exponent $A$ tests $2-2h$. The results of these fits are summarised in Table~\ref{tab:sameS_fitresult} in Appendix~\ref{app:A} are in excellent agreement with the expectations.

Based on the above analysis, we further verify that the UV fixed point is the parafermionic coset $\frac{su(2)_n}{u(1)_n}$ with central charge $\frac{2n-2}{n+2}$ with the perturbing field of conformal weights $h=\bar h = 1-\frac{1}{n}$.

\subsection{Affine $\hat A_n$: shifted $S$-matrix with CDD}

This is an exceptional case, since this TBA system is "on the verge" of having an ultraviolet limit. $L$-functions do not saturate to plateaux they diverge instead, although in a way that physical quantities stay finite. In contrast to the usual plateau case, where e.g. the free energy density reaches its UV value as a power-like function of the dimensionless volume, here $\frac{1}{\log r}$ corrections appear~\cite{AhnBajnok2024}. This is a known phenomenon for e.g. in the sinh-Gordon ~\cite{Zamolodchikov:1992ulx,1996NuPhB.477..577Z,2006JPhA...3912863Z}, Bullough--Dodd and super sinh-Gordon models~\cite{1999NuPhB.556..505A}.

 The TBA system in this case reads as:
\begin{equation}\label{eq:shiftedCDDTBA}
\begin{aligned}
    \epsilon_1(\theta) + (\varphi\star L_{2})(\theta) + (\varphi^{(+\alpha)}\star L_{n})(\theta) &= \frac{r}{2}e^\theta\,, \\
    \epsilon_a(\theta) + (\varphi\star L_{a-1})(\theta) + (\varphi\star L_{a+1})(\theta) &= 0\,, \hspace{2cm} a=2,...,n-1\\
    \epsilon_n(\theta) + (\varphi\star L_{n-1})(\theta) + (\varphi^{(-\alpha)}\star L_{1})(\theta) &= \frac{r}{2}e^{-\theta}\,,
\end{aligned}
\end{equation}
with
\begin{equation}
    \varphi^{(\pm\alpha)} (\theta) = \frac{1}{\cosh(\theta\pm  \alpha)}.
\end{equation}

\begin{figure}
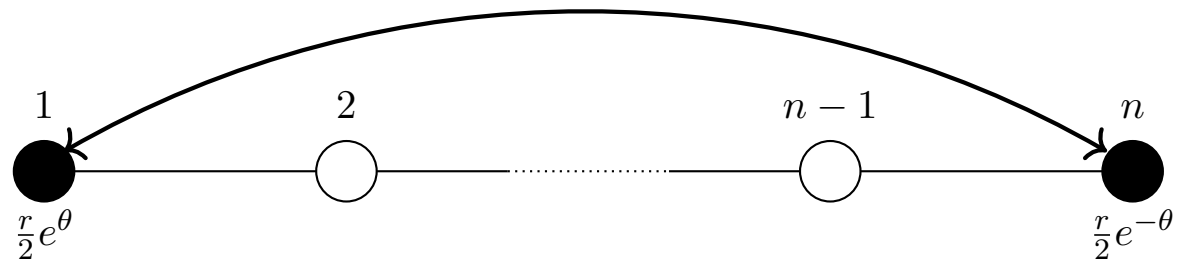

    \centering
            \pgfkeys{/Dynkin diagram,edge length=2cm,fold radius=.5cm}
\begin{DynkinDiagrams}{[labels*={1,2,n-1,n},labels={\frac{r}{2}e^{\theta},,,\frac{r}{2}e^{-\theta}},root radius=0.2cm]A{*o.o*}}\draw[thick,<->] (0.13,0.13) arc (120:60:6.9);\end{DynkinDiagrams}
    \caption{$\hat A_n$ Dynkin diagram for the shifted $S$-matrix with CDD deformation. The top line is thicker and ends in arrows, indicating that the universal kernel is shifted by $\pm\alpha$ for the TBA eq. for $\varepsilon_{1,n}$ respectively.}
    \label{fig:Anhatdiag}
\end{figure}

This TBA system has been studied in~\cite{AhnBajnok2024} in detail both analytically and numerically for $\alpha=0$ and numerically for $\alpha\gg0$. Here we fill the gap and study small but nonzero $\alpha$, where the roaming behaviour is not yet visible. We found that the Liouville reflection function quantisation carries over for non-zero $\alpha$. We also demonstrate the formation of the roaming trajectory as the $\alpha$ parameter is changing. In Section~\ref{sec:IR} we also show this transition is captured in the IR, namely the change in the coefficient of the $T\bar T$ coupling.

In the following, we recall the necessary ingredients from parafermionic Liouville CFT and sinh-Gordon model based on~\cite{AhnBajnok2024}.

\subsubsection{Parafermionic Liouville CFT}

Parafermions~\cite{Zamolodchikov:1985wn} and their perturbations appear in various 1+1 dimensional problems~\cite{Fateev:1991bvBB,Fateev:1991bv}. In particular, the UV limit of the $\hat A_n$ TBA is conjectured to be the parafermionic Liouville conformal field theory (PLFT), with the Lagrangian: 
\begin{equation}
    \mathcal{L}_{\text{PL}_n}=\mathcal{L}_{\text{PF}} + \frac{1}{4\pi}\left(\partial_{\mu}\phi\right)^2-\mu \psi_1 \bar \psi _1 e^{2 b \phi}+\dots\,,
\end{equation}
where $\psi_1$ and $\bar \psi_1$ are $\mathbb Z_n$ parafermions with left and right holomorphic dimensions $h_{\psi}=\bar h_{\bar \psi}=1-1/n$, described by the (formal) Lagrangian $\mathcal{L}_{\text{PF}}$~\cite{1998NuPhB.532..567B}. The PLFT is a conformal field theory with central charge fixed by the parameter $b$ and $n$:
\begin{equation}
    c_{\text{PL}_n}=\frac{3n}{n+2}+6 Q^2\,,\qquad Q=b+\frac{1}{n b}
\end{equation}

Primary fields are vertex operators given by

\begin{equation}
    V_{P}^{(n)}=\sigma_j e^{2\alpha(P)\phi}\,,\qquad \alpha(P)=\frac{Q}{2}+i P\,;\qquad j=0,1,\dots,n-1
\end{equation}
with the "spin fields" $\sigma_j$. The conformal dimension of the vertex operators are
\begin{equation}
    \Delta_{j}=\frac{Q^2}{4}+P^2+\frac{j(n-j)}{2n(n+2)}
\end{equation}

The effective central charge is defined by $c_{\text{eff}} = c-24\Delta_0$ with $\Delta_0$ being the smallest dimension in the theory, given by $j=0$ and $P=0$
\begin{equation}
    c_{\text{eff}}=\frac{3n}{n+2}\,.
\end{equation}

The operators $V_{\pm P}^{(0)}$ are related to each other by the reflection amplitude~\cite{1998NuPhB.532..567B}
\begin{equation}\label{eq:ReflA}
    R^{(n)}_{b}=e^{i\delta^{(n)}(P)} = \left( \frac{\pi \mu \gamma\left(b^2+\frac{1}{n}\right)}{n\left(n^2b^4\right)^{\frac{1}{n}}}\right)^{-\frac{2iP}{b}}\frac{\Gamma\left(1+2iPb\right)\Gamma\left(1+\frac{2iP}{nb}\right)}{\Gamma\left(1-2iPb\right)\Gamma\left(1-\frac{2iP}{nb}\right)}\,.
\end{equation}

\subsubsection{Parafermionic sinh Gordon model: UV expansion from the reflection factors}

The parafermionic sinh-Gordon (PshG) model is given as an integrable perturbation of the PLFT~\cite{1998NuPhB.532..567B,AhnBajnok2024}:
\begin{equation}\label{eq:PshG}
    \mathcal{L}_{\text{PshG}}=\mathcal{L}_{\text{PF}} + \frac{1}{4\pi}\left(\partial_{\mu}\phi\right)^2-\mu \left(\psi_1 \bar \psi _1 e^{2 b \phi}+\eta\psi_1^{\dagger} \bar \psi _1^{\dagger} e^{-2 b \phi}\right)+\dots\,,
\end{equation}
and it has a massive and a massless phase with $\eta=+1$ or $\eta=-1$ respectively. The mass-gap relation has been computed in~\cite{1998NuPhB.532..567B}:
\begin{equation}\label{eq:PshGmassgap}
    \frac{\pi \mu}{n}\gamma(b Q)=\left[\frac{m}{8\Gamma\left(\frac{n+2}{2}\right)}\Gamma\left(1+\frac{n^2b^2}{2(1+nb^2)}\right)\Gamma\left(\frac{n}{2(1+nb^2)}\right)\right]^{2bQ}
\end{equation}

The effective central charge is governed by the primary field with $j=0$ and with the minimum value of the momentum $P$:
\begin{equation}\label{eq:ceffL}
    c^{\text{eff}}(R) = \frac{3n}{n+2}-24P^2+\mathcal{O}(R)\,.
\end{equation}
The Liouville momentum is quantized by the condition (see~\cite{AhnBajnok2024} and references therein):
\begin{equation}\label{eq:LiuQuant}
    \delta^{(n)}(P)=\delta_1 P+\delta_3 P^{3}+\dots=\pi+4 Q P \log x\,;\qquad x=\frac{R}{2\pi}\,.
\end{equation}

The shift parameter $\alpha$ is related to the link between the nodes with source terms in the $\hat{A}_n$ TBA system. In~\cite{AhnBajnok2024}, the quantisation condition was stated for purely imaginary $\alpha$ and numerical comparison has been carried out only for vanishing $\alpha$.

Here we consider non-vanishing real $\alpha$, that can be distributed among the TBA nodes in~\eqref{eq:shiftedCDDTBA} by applying rapidity shifts. In particular, the TBA can be recast in a more symmetric form:
\begin{equation}
    \tilde\epsilon_{i}(\theta)=\frac{r_{\text{eff}}}{2}\left(\delta_{i,1}e^{\theta}+\delta_{i,n}e^{-\theta}\right)-\varphi^{\left(+\frac{\alpha}{n}\right)}\star \tilde L_{i-1}(\theta)-\varphi^{\left(-\frac{\alpha}{n}\right)}\star \tilde L_{i+1}(\theta)\,,\qquad i+n\equiv i\,,
\end{equation}
where the tilde means that the arguments are shifted by $\frac{(2i-n-1)\alpha}{2n}$ and  $r_{\text{eff}}=r e^{-\frac{n-1}{2n}\alpha}$.

Now we are in position to relate $\alpha$ to the Liouville parameter (see~\cite{AhnBajnok2024}):
\begin{equation}\label{eq:alphaid}
    \frac{i\alpha}{n\pi}+\frac{1}{2}=\frac{b}{Q}=\frac{n b^2}{1+nb^2}\,,
\end{equation}
moreover, the effective volume $r_{\text{eff}}$ should be incorporated in~\eqref{eq:PshGmassgap} as $m\rightarrow m e^{-\frac{n-1}{2n}\alpha}$.
Note that this is similar to~\cite{Zamolodchikov:1992ulx}, in choosing the real shifted kernel instead of the imaginary, and can be viewed as the analytic continuation of the $S$-matrix of~\cite{2002PhLB..541..194A,AhnBajnok2024}.
From the above consideration, one can generate the effective central charge as an expansion in $\frac{1}{\log r}$~\cite{AhnBajnok2024}:
\begin{equation}\label{eq:logexp}
    c^{\text{eff}}=\frac{3n}{n+2}+\sum_{k=2}\frac{c^{(n)}_k(\alpha)}{\left(\log r\right)^k} + \mathcal{O}(r)\,.
\end{equation}

Alternatively, the expansion coefficients of the Liouville reflection amplitude given in~\eqref{eq:LiuQuant} can be extracted from the numerical solution to the TBA equation. Our numerical results are collected in Tables~\ref{tab:n2}-\ref{tab:n5} of Appendix~\ref{app:Lrefl} for $n=2,3,4,5$ respectively. The analysis shows, that the Liouville reflection amplitude quantisation survives the analytic continuation and describes the UV limit for finite $\alpha$ as well. Finally, we mention that the $n=2$ case with imaginary $\alpha$ was studied in~\cite{2002PhLB..541..194A}.

\begin{figure}[t]
    \centering
    \includegraphics[width=0.7\linewidth]{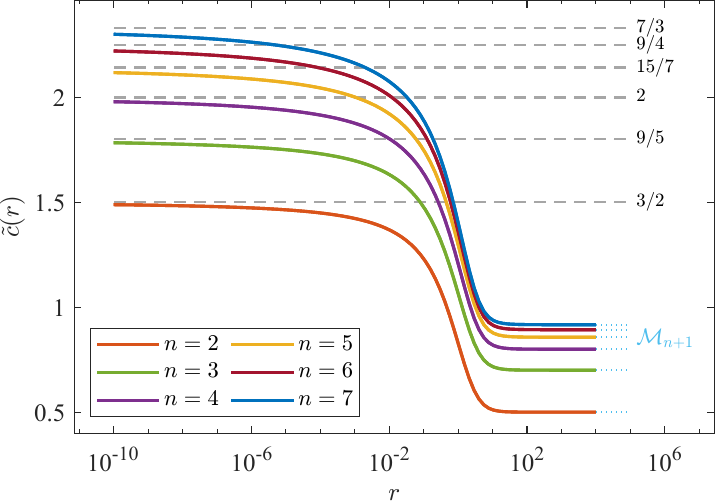}
    \caption{Effective central charges for the shifted $S$-matrix with CDD deformation, with $\alpha=0$.}
    \label{fig:shiftedPlot}
\end{figure}

\begin{figure}[ht]
    \centering
    \includegraphics[width=0.7\linewidth]{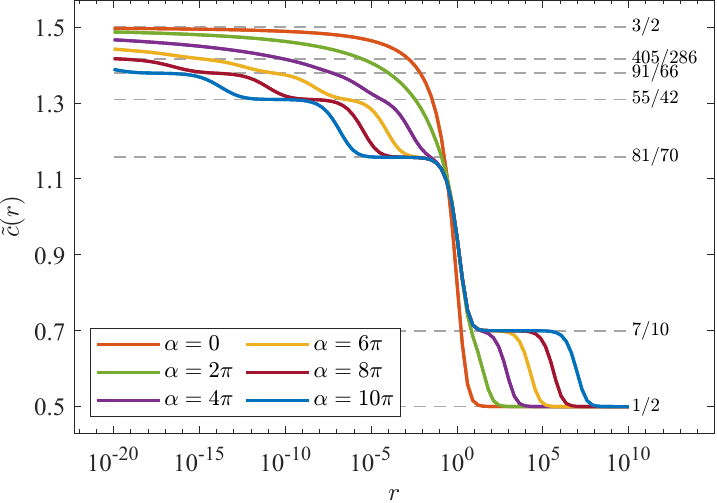}
    \caption{Transition to roaming for $n=2$.
    }
    \label{fig:roaming}
\end{figure}

The volume dependence of the effective central charges is presented for various cases in Figure~\ref{fig:shiftedPlot} and~\ref{fig:roaming}. It is obvious that for $\alpha=0$ the UV central charges are predicted correctly. Increasing $\alpha$ slowly pushes the curves to plateaux predicted in~\cite{AhnBajnok2024} for $n=2$. An interesting feature is that changing $\alpha$ also affects the IR behaviour; see also Section~\ref{sec:hatAIR}.

\section{Non-invertible symmetries and the IR limit}\label{sec:IR}

Motivated by~\cite{2019JHEP...01..026C,TanakaNakayama2024JHEP...11..137N}, there are many other examples of how the non-invertible symmetries play a role along RG flows. See also other recent applications~\cite{2026arXiv260309591B,2025PhRvL.135b1602A,2026arXiv260118667A,2026arXiv260802717A,2026arXiv260824385A}. The minimal criterion is that the residual NIS after perturbing the UV fixed point should be contained in the NIS of the IR theory. Moreover, the set of the topological lines commuting with the perturbing field and their quantum dimensions has to be realised both in the UV and IR perturbed fixed points. We recall these arguments and confront them with the numerical analysis of the TBA in the infrared, which in turn supports the NIS prediction.

Based on~\cite{TanakaNakayama2024JHEP...11..137N} we look for topological defect lines that commute with the UV perturbation. In many cases the set of topological lines matches the set of Verlinde lines, and their properties can be formulated in terms of modular $S$-matrix elements.

The action of a Verlinde line on a primary state is given as:

\begin{equation}\label{eq:topline}
    \mathcal{L}_{n} \ket{m} = \frac{S_{n,m}}{S_{\mathbb I,m}}\ket{m}\,,
\end{equation}
where $\ket{m}=\Phi_m(0)\ket{0}$ is the state corresponding to the primary $\Phi_m$. The general Latin indices run through the primary fields. Later on, they will be denoted by tuples to make everything concrete.

To look for the commuting set of topological lines we recall the criterion of Nakayama and Tanaka and look for a set of lines that satisfy\footnote{The topological lines by definition commute with the chiral spectrum generating algebra~\cite{PetkovaZuber2001PhLB..504..157P}, therefore it is enough to check for primaries.}:
\begin{equation}
    \mathcal{L}_n \Phi_p \ket{m} =  \Phi_p \mathcal{L}_n \ket{m}\,,
\end{equation}
where $\Phi_p$ is the relevant primary inducing the RG flow. Let us define the set of fields $\Lambda_{pm}$, which contains the field in the OPE $\Phi_p\times\Phi_m$. The general condition then reads as:
\begin{equation}
    \sum_{l\in\Lambda_{pm}} \frac{S_{n,l}}{S_{\mathbb I,l}}\ket{l} = \sum_{l\in\Lambda_{pm}} \frac{S_{n,m}}{S_{\mathbb I,m}}\ket{l}\,.
\end{equation}
This condition eventually selects the commuting set of topological lines. This in turn leads to the same results, that recently appeared in a broader context in~\cite{2026arXiv260309591B}.

We note that in a unitary theory, it is sufficient to check the action of the topological line on the perturbing operator to be~\cite{2019JHEP...01..026C}:
\begin{equation}
    \mathcal{L} \Phi_p = d_{\mathcal{L}}\Phi_{p}\,,
\end{equation}
understood from ~\eqref{eq:topline} via the operator state correspondence, and $d_{\mathcal L}=\braket{\mathcal{L}}$ is the quantum dimension of the topological line. In particular, $d_{\mathcal L_n}=\frac{S_{n,\mathbb I}}{S_{\mathbb I,\mathbb I}}$ for Verlinde lines.

\subsection{Double $A_{n-1}$ TBA}

In this case the UV theory is $\frac{su(2)_{n-1}\times su(2)_{n}}{su(2)_{2n-1}}$, with central charge:
\begin{equation}
    c^{n-1,n}_{2n-1} = \frac{3(n-1)n(2n+3)}{(n+1)(n+2)(2n+1)}
\end{equation}
The primary fields $\Phi^{l,l'}_{l''}$  are labeled by a triple of integers 
\begin{equation}
    l=0,1,\dots,n-1,\qquad l'=0,1,\dots,n,\qquad l''=0,1,\dots,2n-1\,.
\end{equation}
subject to the constraint 
\begin{equation}
    l+l'+l'' =0\mod 2\,.
\end{equation}
The chiral conformal weight of these fields are:
\begin{equation}
    h^{l,l'}_{l''} = h_l+h_{l'}-h_{l''}\, \mod 1\,,
\end{equation}
where
\begin{equation}\label{eq:WZWw}
    h_{l} = \frac{\frac{l}{2}\left(\frac{l}{2}+1\right)}{n+2}\,.
\end{equation}
Note that fields can be identified:
\begin{equation}
    \left(l,l',l'' \right) \equiv \left(n-1-l, n-l',2n-1-l''\right)\,.
\end{equation}
The diagonal modular $S$-matrix elements are the following:
\begin{equation}
    S_{(l,l',l''),(k,k',k'')}= S^{(n-1)}_{l,k}S^{(n)}_{l',k'}S^{(2n-1)}_{l'',k''}\,,
\end{equation}
where
\begin{equation}\label{eq:sin}
    S^{(n)}_{ll'} = \sqrt{\frac{2}{n+2}}\sin \left(\frac{\pi (l+1)(l'+1)}{n+2}\right)\,.
\end{equation}
The perturbing field is $(0,0,2)$, and we can apply Nakayama--Tanaka's criterion to get that the set of commuting topological defect lines is given by
\begin{equation*}
    \mathcal{L}^{l,l'}_{0}
\end{equation*}
with all allowed $l,l'$, that is equivalent to the symmetry category $\mathcal L^{(n+1)}_{(r,s)}$ of the diagonal minimal model $\mathcal{M}_{n+1}$. This result has been stated in~\cite{2026arXiv260309591B}.

For the sake of pedagogy, we detail the residual NIS categories in the first few cases, then provide the general picture.

\paragraph{\boldmath$n=2$}

The double $A_1$ flow is the well-known one starting from the tricritical Ising model.
\begin{equation}
\begin{aligned}
    d^{0,0}_{0} &=1\,, &d^{0,2}_{0} &= 1\,, & d^{1,1}_{0} &= \sqrt{2}\,.\\
    \end{aligned}
\end{equation}
These are the total set of quantum dimensions of the Ising model. From the coset fusion rules, we have the nontrivial fusions:
\begin{equation}
    \begin{aligned}
            \mathcal L_{0}^{0,0}\times \mathcal L_{0}^{0,0} &= \mathcal L_{0}^{0,0}\\
            \mathcal L_{0}^{0,2}\times \mathcal L_{0}^{0,2} &= \mathcal L_{0}^{0,0}\\
            \mathcal L_{0}^{1,1}\times \mathcal L_{0}^{0,2} &= \mathcal L_{0}^{0,2}\times \mathcal L_{0}^{1,1}=\mathcal L_{0}^{1,1}\\
            \mathcal L_{0}^{1,1}\times \mathcal L_{0}^{1,1} &= \mathcal L_{0}^{0,0} + \mathcal L_{0}^{0,2}
    \end{aligned}
\end{equation}
which is equivalent to the Ising fusion category ($\mathbb Z_2$ Tambara--Yamagami category), with the identification $\mathcal{L}^{0,0}_0=\mathbb I$ is the identity line, $\mathcal L_{0}^{0,2}=\eta$, the invertible line and $\mathcal{L}_0^{1,1}=\mathcal N$ the non-invertible line.

\paragraph{\boldmath$n=3$}

The commuting defect lines have the following quantum dimensions:
\begin{equation}
\begin{aligned}
    d^{0,0}_{0} &=1\,, &d^{2,0}_{0} & = 1\,, &  d^{0,2}_{0} &= \varphi_G\,, & d^{2,2}_{0} &= \varphi_G\,,&d^{1,3}_{0} &= \sqrt{2}\,,&  d^{1,1}_{0} &= \sqrt{2} \varphi_G\,,\\
    \end{aligned}
\end{equation}
where $\varphi_G=\frac{1+\sqrt{5}}{2}$ is the golden ratio. This is the same as the full list of quantum dimensions of the tri-critical Ising model. The non-trivial fusions are collected in Table~\ref{tab:TrisingFusion}.

\begin{table}[]
    \centering
    \begin{tabular}{c|ccccc}
        $\times$ & $\mathcal{L}_{0}^{2,0}$ &  $\mathcal{L}_{0}^{0,2}$&  $\mathcal{L}_{0}^{2,2}$ &  $\mathcal{L}_{0}^{1,3}$ & $\mathcal{L}_{0}^{1,1}$\\\hline\\[-1em]
         $\mathcal{L}_{0}^{2,0}$ & $\mathcal{L}_{0}^{0,0}$ & $\mathcal{L}_{0}^{2,2}$ & $\mathcal{L}_{0}^{0,2}$ & $\mathcal{L}_{0}^{1,3}$ & $\mathcal{L}_{0}^{1,1}$\\ \\[-1em]
         $\mathcal{L}_{0}^{0,2}$ & $\mathcal{L}_{0}^{2,2}$ & $\mathcal{L}_{0}^{0,0}$ + $\mathcal{L}_{0}^{0,2}$ & $\mathcal{L}_{0}^{2,2}$ + $\mathcal{L}_{0}^{2,0}$ &
         $\mathcal{L}_{0}^{1,1}$ & $\mathcal{L}_{0}^{1,1}$ + $\mathcal{L}_{0}^{1,3}$ \\ \\[-1em]
         $\mathcal{L}_{0}^{2,2}$ & $\mathcal{L}_{0}^{2,0}$ &
         $\mathcal{L}_{0}^{2,2}$ + $\mathcal{L}_{0}^{0,2}$ & $\mathcal{L}_{0}^{0,0}$ + $\mathcal{L}_{0}^{0,2}$ & $\mathcal{L}_{0}^{1,1}$ &
         $\mathcal{L}_{0}^{1,1}$ + $\mathcal{L}_{0}^{1,3}$ \\ \\[-1em]
         $\mathcal{L}_{0}^{1,3}$ & $\mathcal{L}_{0}^{1,3}$ &
         $\mathcal{L}_{0}^{1,1}$ & $\mathcal{L}_{0}^{1,1}$ & $\mathcal{L}_{0}^{0,0}$ + $\mathcal{L}_{0}^{2,0}$ & $\mathcal{L}_{0}^{0,2}$ + $\mathcal{L}_{0}^{2,2}$ \\ \\[-1em]
         $\mathcal{L}_{0}^{1,1}$ & $\mathcal{L}_{0}^{1,1}$ &
         $\mathcal{L}_{0}^{1,1}$ + $\mathcal{L}_{0}^{1,3}$ &
         $\mathcal{L}_{0}^{1,1}$ + $\mathcal{L}_{0}^{1,3}$ &
         $\mathcal{L}_{0}^{0,2}$ + $\mathcal{L}_{0}^{2,2}$ &
         $\mathcal{L}_{0}^{0,0}$ + $\mathcal{L}_{0}^{0,2}$ + $\mathcal{L}_{0}^{2,0}$ + $\mathcal{L}_{0}^{2,2}$ 
    \end{tabular}
    \caption{Fusion rules of the residual NIS category in the case of the double $A_2$ type flow. The fusion rules are equivalent to the ones of the tricritical Ising model given in~\eqref{eq:trisinglines}.}
    \label{tab:TrisingFusion}
\end{table}

These are the same as the topological defect lines in the tricritical Ising model with the following identification:
\begin{equation}\label{eq:trisinglines}
    \begin{aligned}
        \mathcal{L}_{0}^{0,0} &= \mathbb I\,, & \mathcal{L}_{0}^{2,0} &= \eta\,, & \mathcal{L}_{0}^{0,2} &= \mathcal{W}\,,&\mathcal{L}_{0}^{2,2}&=\eta\mathcal W\,,&\mathcal{L}_{0}^{1,3}&=\mathcal N\,,& \mathcal{L}_{0}^{1,1}&=\mathcal{NW}\,.\\
    \end{aligned}
\end{equation}

\paragraph{General \boldmath$n$}
The above pattern continues, and in any case, the residual non-invertible symmetry coincides with the one of the infrared theory.
 Up to normalisation, the modular $S$-matrix elements corresponding to the residual NIS are the following:
\begin{equation}
    \tilde{S}_{(l,m)(l',m')} \propto \sin \left( \frac{\pi(l+1)(l'+1)}{n+1}\right) \sin \left( \frac{\pi(m+1)(m'+1)}{n+2}\right)\,.
\end{equation}
Let $p=n+1\,;\, r=l+1=1,2,\dots p-1$ and $s=m+1=1,2,\dots,p$ and remind the reader that both $l+m$ and $l'+m'$ are even. Now, the above formula for the $S$-matrix elements are the same as the modular $S$-matrix for the diagonal modular invariant of the $\mathcal{M}_{n+1,n+2}$ minimal model\footnote{Restricting the Kac-table to the field with $r+s$ even selects the primaries without redundancy}.

This leads to the conclusion, that the family of irrelevant operators generating the flow is the generalised $T \bar T$ family. This was already speculated in~\cite{Zamolodchikov:1991vg} and it is also reflected in our numerics. We fitted the following functional form in the IR region $200\leq r \leq 2000$ of the TBA equations
\begin{equation}
    \tilde{c}(r) = c(n) +\sum_{k=1,2,3} \tilde{c}_k r^{-2k}\,,
\end{equation}
where $c(n)=1-\frac{6}{(n+1)(n+2)}$. The results for the first two coefficients for $n=3,\dots,7$ are collected in Table~\ref{tab:identityIRfits}. It is then clear that this flow enters from the direction of $T \bar T$, consistent with the expectation regarding pure CDD factors.
In the first two orders in $r^{-2}$, contributions are solely generated by $T \bar T$, and their contribution are given in terms of Catalan numbers~\cite{Zamolodchikov:1991vh,2020PhRvD.101b6011G,2024arXiv240806729B}, in particular, for the CDD factor~\eqref{eq:CDDfactor} they are expected to be:
\begin{equation}\label{eq:catalan}
    \begin{aligned}
        \tilde c_1 &= c(n)^2 \frac{2\pi}{3} e^{\alpha}\,, \\
        \tilde c_2 &= 2 c(n)^3 \left(\frac{2\pi}{3}e^{\alpha}\right)^2\,.
    \end{aligned}
\end{equation}

It is worth mentioning that the sign of $~\tilde c_1$ corresponds to a $T\bar T$ deformation leading to a square root singularity~\cite{2016JHEP...10..112C,SmirnovZamolodchikov2017NuPhB.915..363S} and the presence of the full odd-$s$ tower is needed to render the theory UV-complete, see also~\cite{2022JHEP...08..179A,2021JHEP...10..062C}.

This also shows the power of non-invertible symmetries. Using the residual NIS and its relation to the category of the infrared theory, one can identify pure generalised $T \bar T$ flows. Indeed, this TBA was obtained by the pure CDD deformation of the massless IR scattering theory, as expected~\cite{SmirnovZamolodchikov2017NuPhB.915..363S}.

\begin{table}[h]
    \centering 
    \begin{tabular}{|c|c|c|c|c|c|} \hline 
        $n$ & $3$ & $4$ & $5$ & $6$ & $7$\\ \hline\hline 
        $\tilde{c}_1$ & \footnotesize  $1.02625359983(3)$ & \footnotesize $1.34041286516(4)$ & \footnotesize $1.53873925804(6)$ & \footnotesize $1.6696389511(1)$ &\footnotesize  $1.759873650(1)$ \\ \hline 
        $\tilde{c}_2$ & $3.009170(5)$ & $4.491807(5)$ & $5.524770(9)$ & $6.24462(3)$ & $6.75771(3)$\\ \hline  
    \end{tabular} 
    \caption{Fit results in the identity deformation (double $A_{n-1}$ type TBA). The coefficients $\tilde c_1$ and $\tilde c_2$ are in excellent agreement with the theoretical prediction for a pure $T\bar T$ perturbation}
    \label{tab:identityIRfits} 
\end{table}

\subsection{$D_n$ TBA system}\label{sec:dnIR}

This model has $\frac{su(2)_n}{u(1)_n}$ as the ultraviolet limit. The central charge of this theory is:
\begin{equation}
    c^n_{\text{PF}} = \frac{2n-2}{n+2}\,.
\end{equation}
The primary fields are labeled by a pair of integers $(l,m)$
\begin{equation}
    l=0,1,\dots,n;\qquad m\in\mathbb Z_{2n}\,.
\end{equation}
satisfying the condition
\begin{equation}
    l + m \equiv0\mod 2\,.
\end{equation}
Such a primary has the conformal weight:
\begin{equation}
    h_{l,m}=\frac{l(l+2)}{4(n+2)}-\frac{m^2}{4n} \mod \mathbb Z\,,
\end{equation}
and the modular $S$-matrix elements are given as:
\begin{equation}
    S_{(l,m),{(l',m')}} = \sqrt{\frac{2}{n}}S^{(n)}_{(l,l')} e^{\frac{i\pi m m'}{n}}\,.
\end{equation}

In the UV parafermionic description, the perturbing operator is a linear combination of $\Phi_{(0,2)}$ and $\Phi_{(0,-2)}$~\cite{Fateev:1991bv,2026JHEP...09..123Z}. Both commute with the Verlinde lines $\mathcal{L}_{(l,0)}$, with all allowed i.e. even $l$, which coincides with defect lines that form the $\mathcal L^{(n+1)}_{(1,s|\text{odd})}$ subcategory of the commutant of $\Phi_{3,1}$ $\mathcal L^{(n+1)}_{(1,s)}$ in the IR fixed point~\cite{2026arXiv260309591B,2026JHEP...09..123Z}. Considering these flows from the IR point of view, they are expected to enter from the direction of $\Phi_{3,1}$ together with possible generalised $T \bar T$~\cite{Fateev:1991bv,Feveratietal1996PhLB..374...64F}, with the opposite sign of the $\Phi_{3,1}$ coupling, compared to the Zamolodchikov case cf. Section~\ref{sec:masslessAn}. From the IR perspective, there is no apparent reason for the residual category to be smaller than the commutant of $\Phi_{3,1}$; whether the full commutant is realised is decided in the UV. Apparently,~\cite{2026arXiv260309591B} considers flows generated purely by $\Phi_{(0,2)}$, while ~\cite{2019JHEP...01..026C} and ~\cite{2026JHEP...09..123Z} recognise the importance of the proper linear combination (as conjectured in~\cite{Fateev:1991bv}):
\begin{equation}
\begin{aligned}
    \mathcal{S}_{D_n} &= \mathcal{S}_{su(2)_n/u(1)_n} - \lambda\int d^2 x V^{(\pi)}_{(0,\pm2)}\,.\\
    V^{(\pi)}_{(0,\pm2)}&=e^{\frac{i \pi}{n}}\Phi_{(0,2)}+e^{\frac{-i \pi}{n}}\Phi_{(0,-2)}\,.
\end{aligned}
\end{equation}
However,~\cite{2026JHEP...09..123Z} only considers its commutant with the Verlinde lines and~\cite{2019JHEP...01..026C} proposes that Wilson lines surviving the coset construction provide the symmetry category.

Here we clarify the UV commutant using only the coset objects. Recall, that the action of a Verlinde line on $\Phi_{0,\pm2}$ is 
\begin{equation}
    \mathcal{L}_{(l,m)}\Phi_{(0,\pm2)} = d_l e^{\pm \frac{2 i\pi m}{n}} \Phi_{(0,\pm2)}\,.
\end{equation}

Note, that the parafermionic theory has a charge conjugation line $\mathcal{C}$ acting as $\mathcal{C}\Phi_{(l,m)}=\Phi_{(l,-m)}$, which maps the two perturbing fields to each other. This line is invertible, since $\mathcal{C}^2=1$, meaning that its quantum dimension is $1$.

One can introduce $\mathcal{C}$-twisted lines $\mathcal{N}_{(l,m)}=\mathcal{C}\circ\mathcal{L}_{(l,m)}$, with the specific action:
\begin{equation}
    \mathcal{N}_{(l,m)}\Phi_{(0,\pm2)} = d_l e^{\pm \frac{2 i\pi m}{n}} \Phi_{(0,\mp2)}\,.
\end{equation}

It is easy to check that the perturbation $V^{(\pi)}_{(0,\pm2)}$ commutes with the twisted lines whenever $2+2m=0 \mod 2n$, that is $m=-1,n-1$. The two can be related by the identification $(l,m)\sim (n-l,m+n)$, so we can safely choose $-1$. This then picks odd $l$s, for the commuting twisted lines. With the addition of the twisted lines that are topological but not Verlinde, we recover the full commutant of $\Phi_{3,1}$, with the expected fusion in the infrared. Note that an equivalent set of lines was introduced in~\cite{2019JHEP...01..026C} as Wilson lines of the WZW theory that remain in the coset, while we use a direct coset realisation.

This addition has a natural interpretation, at least for $n=4$. The reduced subcategory only has integer quantum dimensions, but the two additional ones contain the irrational $\sqrt{3}$, meaning that the flow cannot be trivially gapped~\cite{2019JHEP...01..026C}. This pairs well with the assumption that it describes a massless flow.

Note that this category still allows for additional generalised $T \bar T$ deformations in the IR along $\Phi_{3,1}$. Actually, for these models, the coupling of $T\bar T$ (at least for even $n$), is conjectured to be  vanishing~\cite{Feveratietal1996PhLB..374...64F}. The IR fits presented in Table~\ref{tab:IRfits} suggest that it could be also the case for odd $n$s. The $r^{-2}$ and $r^{-2}\log r$ terms are generated via the resonance between $\Phi_{3,1}$ and $T\bar T$. The $r^{-2}$ the coefficients are $+0.206042$ and $-0.206557$ for $A_5$ and $D_5$ respectively while the $r^{-2}\log r$ coefficients are $-0.3267128$ vs. $+0.326777$, the sign flip this suggest these are generated purely by $g^3$, while for $A_7$ and $D_7$ they are $-0.1605$ and $-0.182$ attributed to $g^4$, and the $r^{-2}\log r$ coefficients are $0.27852$ vs. $0.281973$. Moreover, the addition of one more CDD factor (see sections~\ref{sec:hatDIR} and~\ref{sec:hatAIR} below and Table~\ref{tab:IRfits} in Appendix~\ref{app:C}) affects only the $r^{-2}$ coefficient significantly, but not the $r^{-2}\log r$. This is consistent with, but does not prove, the absence of an independent $T \bar T$ term. It would be interesting to show this rigorously, but this is out of the scope of the present paper.

\subsection{$\hat D_n$ TBA system}\label{sec:hatDIR}

It is still possible to add more CDD factors to the $S$-matrix: an extra CDD factor leads to introducing a universal kernel between the massless modes, i.e. it would change the first two lines in eq.~\eqref{eq:sameSTBA} to
\begin{equation}\label{eq:Dhat}
    \begin{aligned}
        \epsilon_1(\theta) + (\varphi \star L_2)(\theta)+ (\varphi \star L_n)(\theta) &= \frac{r}{2}e^\theta \\
        \epsilon_n(\theta) + (\varphi \star L_2)(\theta)+ (\varphi \star L_1)(\theta) &= \frac{r}{2}e^{-\theta} \\
    \end{aligned}
\end{equation}
However, we numerically found that this has no UV completion in the usual sense; instead, it shows a Hagedorn-like singularity: the iteration of the TBA equation stops converging and the effective central charge blows up as $r\rightarrow r_c$. On the other hand, this extra CDD addition produces generalised $T\bar T$ terms in the IR. In particular, it increases the coupling of $T\bar T$ by the same amount as in the pure CDD deformation case. The supporting data is presented in Table~\ref{tab:CDDshift}.

\subsection{Massless $A_n$ TBA system}\label{sec:masslessAn}

For the sake of completeness, we recall the result of Nakayama and Tanaka, that states that in the Zamolodchikov flows between minimal models $\mathcal{M}_{(n+2)}$ and $\mathcal{M}_{(n+1)}$ the residual NIS category is again $\mathcal L^{(n+1)}_{(1,s)}$, which is compatible with $\Phi_{3,1}$ irrelevant perturbation in the IR.

We remind the reader that the $A_n$ and $D_n$ models have been studied numerically in~\cite{Feveratietal1996PhLB..374...64F}. Indeed, the authors found that these flows were generated by the irrelevant operator $\Phi_{3,1}$ with the opposite sign of the coupling.

The results apparently show an interesting sign pattern depending on $n$: the coefficient of $r^{-2}$, related to $T\bar T$ is compatible with $\sigma_n c(n)^2\frac{2\pi}{3}=-\cos \left(\frac{n \pi}{2}\right)c(n)^2\frac{2\pi}{3}$ (cf. eq.~\eqref{eq:catalan})\footnote{It would be interesting to understand this pattern from the $S$-matrix deformation, but this is out of the scope of the present paper. See also the discussion in Section~\ref{sec:outro}}.

\subsection{$\hat{A}_n$ TBA system}\label{sec:hatAIR}

In this case the UV theory is a parafermionic Liouville theory and the perturbing operator is
\begin{equation}
    \psi_1^\dagger \bar \psi_1^\dagger e^{-2b\phi}\,.
\end{equation}
This field naturally inherits the parafermionic properties from the previous case: on the parafermionic field factor $\psi_1^\dagger \bar \psi_1^\dagger\sim\Phi_{(0,-2)}$ the Verlinde lines act as in Section ~\ref{sec:dnIR}, while the Liouville vertex operator $e^{-2b\phi}$ is transparent. Therefore, the residual NIS is expected to be the same $\mathcal L^{(n+1)}_{(1,s|\text{odd})}$ as in the naive treatment of $D_n$ flows above. From the infrared point of view, this model is an additional CDD factor to the massless $A_n$, therefore one expects additional generalised $T\bar T$ deformations only, commuting with the topological lines. Therefore, one does not expect the category of $A_n$ to shrink. Below we present the numerical argument that indeed, the deformation only adds generalised $T\bar T$ terms and sketch how the NIS category could survive.

Let us start with $n=2$. This case is exceptional, since there is no irrelevant $\Phi_{3,1}$ in the infrared theory (Ising CFT). We can still compare the infrared behaviour of the effective central charge from Zamolodchikov's TBA and the deformed one:
\begin{equation}\label{eq:a2IRs}
\begin{aligned}
    \tilde{c}(r)_{A_2}&\approx\frac{1}{2} + 0.52359877562(2) r^{-2} +1.0966188(3)r^{-4}+\dots\,,\\
    \tilde{c}(r)_{\hat A_2}&\approx\frac{1}{2} + 1.047197549(2) r^{-2} +4.386521(9)r^{-4}+\dots\,.
\end{aligned}
\end{equation}
This shows that the coefficient of $T\bar T$ is doubled, as it is expected by adding an extra CDD factor, i.e. by changing $\sigma_n\rightarrow\sigma_n+1=2$ for $n=2$. While the deformation is generated by the $T\bar T$ family, the naive symmetry category is shrunk from the Ising category to its invertible part.

To the best of our knowledge, not much is known about the full set of topological lines in the parafermionic Liouville theories. However, at least for $n=2$ we can find a topological line similar to the $D_n$ case. Let us consider the two terms in~\eqref{eq:PshG} together 
\begin{equation}
    V_{\eta,b}=\psi_1 \bar \psi _1 e^{2 b \phi}+\eta\psi_1^{\dagger} \bar \psi _1^{\dagger} e^{-2 b \phi}\,
\end{equation}
as a perturbation of the parafermion plus a massless boson.
For $n=2$, $\psi_1=\psi_1^\dagger$ is the "Ising Majorana fermion"\footnote{For $n=2$ the theory is the Ising model coupled to a massless free boson.}, and the massless perturbation ($\eta=-1$) has the form $\psi \bar \psi \sinh 2b\phi$, while the massive ($\eta=1$) is $\psi \bar \psi \cosh 2b\phi$. We are in a similar situation as in the $D_n$ case above: we can combine the non-invertible Ising line $\mathcal N$ with the reflection $\mathcal{R}_{\phi}$ that maps $\phi$ to $-\phi$ to $\mathcal N \otimes \mathcal{R}_{\phi}$. Since $\mathcal{N}$ acts as a sign on $\psi\bar \psi$, this combined line commutes with the massless perturbation. In turn, this means that there is an additional commuting line with quantum dimension $\sqrt{2}$ (inherited from $\mathcal N$) in the massless case, creating room for the full Ising category. Note that this line does not commute with the massive perturbation, and the corresponding quantum dimensions are only integers, consistent with~\cite{2019JHEP...01..026C}.

In Table~\ref{tab:IRfits} in Appendix~\ref{app:C} we present the results of IR fits for $n=4,5,6,7,8$, similarly to~\cite{Feveratietal1996PhLB..374...64F}. The differences between $r^{-2}$ coefficients are collected in Table~\ref{tab:CDDshift} and are in excellent agreement with an additional CDD factor (cf. eq.~\eqref{eq:catalan}) for small $n$. The difference is still compatible, but gradually degrades for larger values of $n$. This clearly shows that indeed, the extra CDD type deformation is caused by a generalised $T\bar T$ deformation.

\begin{table}[t]
\centering
\begin{tabular}{|c|c|c|c|}
\hline
$n$ & $c(n)^2\tfrac{2\pi}{3}$ & $\hat A_n - A_n$  & $\hat D_n - D_n$ \\
\hline \hline
$4$ & $1.34041$ & $1.340581(2)$  & $1.340623(3)$\\
$5$ & $1.53874$ & $1.536538(9)$  & $1.53530(3)$ \\
$6$ & $1.66964$ & $1.6511(5)$    & $1.6252(6)$  \\
$7$ & $1.75987$ & $1.7330(2)$    & $1.654(1)$   \\
$8$ & $1.82445$ & $1.7695(8)$    & $1.584(2)$   \\
\hline
\end{tabular}
\caption{Difference between the $r^{-2}$ coefficient of $\tilde c(r)$ produced by the additional CDD factor
($\hat A_n$ vs.\ $A_n$ and $\hat D_n$ vs.\ $D_n$, from Table~\ref{tab:IRfits}) compared to the
pure $T\bar T$ prediction in eq.~\eqref{eq:catalan} at $\alpha=0$, with errors of the fits.}
\label{tab:CDDshift}
\end{table}

The data shows, that the extra CDD is consistent with the idea of generalised $T\bar T$ for any $n$. However, the simple $n=2$ argument does not work for general $n$: the two parafermion bilinears gather different phase factors under the action of the corresponding twisted line as in the $D_n$ case. Using the notation there: $\Phi_{(0,2)}=\psi_1\bar \psi_1$ and $\Phi_{(0,-2)}=\psi^\dagger_1\bar \psi^\dagger_1$:
\begin{equation}
    \mathcal{N}_{(l,m)}\otimes R_{\phi} \Phi_{(0,\pm2)} e^{\pm 2b\phi} = d_l e^{\pm\frac{2i\pi m}{n}}\Phi_{(0,\mp2)} e^{\mp 2b\phi}\,.
\end{equation}
To have a commuting line, the resulting phase should be $-1$, which holds only for $n\equiv 2\, \text{mod}\,4$; therefore, the $n=2$ argument does not generalise. There are two possible ways out: the actual massless ray is not $\eta=-1$, but there is a relative phase, $e^{-\frac{2 i \pi}{n}}$ between the terms, as in the $D_n$ case; or the parafermionic Liouville or the parafermion$\otimes$boson theory has more combined topological lines. The complete treatment is out of the scope of the present paper.

We also studied the effect of non-vanishing $\alpha$. For $n=4,5$ we collected the $r^{-2}$ coefficients in Table~\ref{tab:alphaTTbar}, together with the expected value from the form of the CDD deformation. Our results show that the CDD factor results in the expected modification of the $T\bar T$ coupling.

\begin{table}[t]
\centering
\begin{tabular}{|c|cc|cc|}
\hline
 & \multicolumn{2}{c|}{$n=4$ }
 & \multicolumn{2}{c|}{$n=5$ } \\
$\alpha$ & $\tilde c_{T\bar T}(\alpha)$ & $c^2\tfrac{2\pi}{3}(e^{\alpha}-1)$
 & $\tilde c_{T\bar T}(\alpha)-\tilde c_{T\bar T}(0)$ & $c^2\tfrac{2\pi}{3}(e^{\alpha}-1)$ \\
\hline\hline
$0$    & $-0.000032(8)$ & $0$   & $0$    & $0$  \\
$0.25$ & $0.380680(8)$  & $0.380711$ & $0.4371(3)$   & $0.437041$  \\
$0.5$  & $0.869522(9)$  & $0.869554$    & $0.9983(3)$   & $0.998213$   \\
$2\pi$ & $716.4397(7)$  & $716.4395$   & $822.4485(2)$ & $822.4433$ \\
\hline
\end{tabular}
\caption{$\alpha$-dependence of the $r^{-2}$ coefficient $\tilde c_{T\bar T}$ of $\tilde c(r)$ in the $\hat A_n$
TBA (fits with fixed exponents in $r\in[10^4,10^7]$, Tables~\ref{tab:An1} and~\ref{tab:An2}) compared
with the $T\bar T$ prediction~\eqref{eq:catalan}. For $n=4$ the $\alpha=0$ value vanishes because the
massless $A_4$ flow contributes $\sigma_4\, c^2\tfrac{2\pi}{3}$ with $\sigma_4=-1$, so
$\tilde c_{T\bar T}(\alpha)=c^2\tfrac{2\pi}{3}(e^{\alpha}-1)$; for $n=5$ the $\alpha=0$ value
$1.7352(2)$ is the resonant $g^3$ companion of the $r^{-2}\log r$ term, so only differences are universal.}
\label{tab:alphaTTbar}
\end{table}

\subsection{IR limit summary}

Let us summarise the features of the massless RSOS flows from the IR perspective. The perturbations allowed by the non-invertible symmetries are $T\bar T$ and $\Phi_{3,1}$, of chiral dimensions $h_T=2$ and $h_{3,1}^{n+1}=1+\frac{2}{n+1}$, except for $n=2$, where $\Phi_{3,1}$ does not exist in the IR CFT. For general $n$, one can write the formal IR action:

\begin{equation}
    \mathcal{S}^n_{\text{RSOS}} = \mathcal{S}_{\mathcal{M}_{n+1}} + \int d^2x\left(g\Phi_{3,1}(x)+\sum_{s\geq1} \beta_s(T\bar T)_s(x)\right)
\end{equation}

The effective central charge can be written in a formal perturbative series, in terms of the couplings and correlation functions of $T\bar T$ and $\Phi_{3,1}$~\cite{Feveratietal1996PhLB..374...64F}. Terms originating from $\Phi_{3,1}$ organise as powers of $r^{-\frac{4}{n+1}}$, corresponding to powers of $g$. The first nonvanishing term is due to $g^2$, while the third contains information on the sign of $g$. We reproduce the finding of~\cite{1993NuPhB.400..547K,Feveratietal1996PhLB..374...64F}: the $A$ and $D$ type diagrams correspond to opposite signs of $g$. Moreover, we also conjecture based on the numeric findings and the construction, that the $D$ type theories (pure RSOS S-matrix deformations) are pure $\Phi_{3,1}$ perturbations, while $A$ type (Zamolodchikov flows, shifted $S$-matrix deformations) correspond to the combination of $\Phi_{3,1}$ and the $T\bar T$ family, with an interesting sign pattern of the $T\bar T$ coupling. Adding more CDD factors ($\hat A, \hat D$) introduces additional coupling to $T\bar T$ and keeps $g$ untouched, compatible with the expectation that CDD deformation does not affect the matrix structure of the scattering theory.

Terms originating from $T\bar T$ are powers of $r^{-2}$ which is the contribution of $T\bar T$ itself, while the next member of the family contributes starting from $r^{-6}$. For pure CDD deformations, the $r^{-2}$ coefficients are compatible with the expectation from the expansion of the CDD factor, as is the change in the coefficients when adding one more CDD factor.

Using the form of the UV perturbation, we obtained the NIS category along the flows. In some cases, this has been done completely, while in others only partially. In particular, for the $D_n$ type TBA systems we showed that one has to include topological, but non-Verlinde lines. In all cases the IR perturbation is compatible with symmetry category obtained in the UV.

\section{Conclusions and outlook}\label{sec:outro}

In this work, we studied massless deformations of RSOS scattering theories. Although these models were previously conjectured~\cite{AhnBajnok2024}, no detailed systematic study has been worked out. We carried out an in-depth analysis both in the UV and IR, comparing the various deformations. The main features are collected in Table~\ref{tab:models}.

The possible UV finite deformations are limited: adding "too much" CDD leads to UV incomplete theories. The "maximal" deformation of the massless RSOS is the $\hat A_n$ TBA system, that is still UV complete, but already develops certain infinities in the UV, leading to logarithmic corrections in the effective central charge. This usually signals the UV fixed point is not governed by a rational CFT. Indeed, it was conjectured in~\cite{AhnBajnok2024} that the $\hat A_n$ TBA system has a parafermionic Liouville theory as the UV fixed point. Moreover, for vanishing shift parameter, the authors verified numerically that the logarithmic corrections can be determined from the corresponding Liouville reflection amplitude. In this work, we demonstrated this for various values of non-vanishing $\alpha$.

Other deformations studied in this work lead to UV complete theories, with "usual", finite plateaux TBA systems and rational CFTs as UV fixed points.

However, in the IR scattering we can add more CDD factors, and further analyse its effect to the couplings using numerical fits to the result of TBA equations in the convergent IR regime.

We summarised the UV and IR features of the models in the context of non-invertible symmetries. It is clear that the pure CDD-type deformations lead to generalised $T\bar T$-type deformations in the IR, while introducing an RSOS-type non-diagonal massless deformation corresponds to the inclusion of the $\Phi_{3,1}$ irrelevant operator. Indeed, we confirmed this observation with a detailed numerical account of the infrared limit of the TBA equations together with the residual NIS category from the UV point of view. Numerically solving the TBA equations in the deep IR, we were able to extract some features of the IR theory explicitly: scalar CDD deformations lead to generalised $T\bar T$ deformations with the total symmetry category of the IR fixed point as the NIS category along the flow; deformations with an RSOS structure are only compatible with the inclusion of the $\Phi_{3,1}$ perturbation with its commutant $\mathcal{L}^{(n+1)}_{(1,s)}$ as the residual NIS. To recover the correct symmetry category, it is extremely important to \emph{i)} identify the exact form of the UV perturbation and \emph{ii)} find all commuting topological lines when they are not exhausted by the Verlinde lines. Additional CDD factors do not affect the coupling of $\Phi_{3,1}$, but change the $T\bar T$ coupling by the \emph{same} amount as the pure CDD factor case. For the double $A_{n-1}$ and $\hat A_2$, we show that CDD factors leave the symmetry category unaffected and argue that this holds in general.

Along the way, we noticed several numerical observations that deserve better understanding in the future. It would be interesting to prove or disprove that $D_n$ is indeed a \emph{pure} $\Phi_{3,1}$ perturbation without generalised $T\bar T$, along with the conjectured $T\bar T$ coupling sign pattern in the $A_n$ TBA systems. It would also be interesting to derive the RSOS scattering structure from the presence of $\Phi_{3,1}$ and understand in detail how the different sign of its coupling leads to different shifts in the argument of the RL and LR $S$-matrices.

\acknowledgments

The authors are grateful to Zoltán Bajnok and Changrim Ahn for various discussions. ML was supported by the Bolyai János Research Scholarship of the Hungarian Academy of Sciences. KST acknowledges funding from the grant NKKP STARTING-152796.

\paragraph{AI Disclaimer} Different models of Anthropic's Claude and OpenAI's ChatGPT were used for literature search and for various revision cycles of the manuscript. References were checked for relevance, while suggested changes were considered and implemented by the authors. In particular, these tools motivated and assisted the inclusion of the non-trivial topological lines and the correct treatment of the reflection amplitude for non-zero $\alpha$.

\appendix

\section{UV limits}\label{app:A}

In this appendix we present the results of numerical fits to the effective central charge in the UV limit of the double $A_{n-1}$ and $D_n$ TBA systems, i.e. the identity and the same $S$-matrix deformations, respectively. The forms of the fit functions are given under~\eqref{eq:IdUV} and~\eqref{eq:sameS_fitfunc}. In both cases $A=2-2h$, while $\mathcal B$ of~\eqref{eq:IdUV} is the coefficient of $r^2$ and $B$ of~\eqref{eq:sameS_fitfunc} is the coefficient of the $r^2 \log r$ term. 

In the double $A_{n-1}$ case $\tilde c_1$ is expected to be zero. For $D_n$ with $n$ even, odd terms are expected to be zero, while for odd $n$ the odd terms with $k < n$ vanish, the first allowed one, $k = n$, being the $r^2$ bulk term.

\begin{table}[h]
\centering
\begin{tabular}{|c|c|c|}\hline
 & $n=3$ & $n=4$ \\ \hline\hline
$A$           & $0.5714285705(3)$       & $0.4444444450(4)$ \\ \hline
$\mathcal B$  & --                      & $0.47754(7)$ \\ \hline
$\tilde c_1$  & $7(12)\cdot 10^{-13}$   & $-5(2)\cdot 10^{-12}$ \\ \hline
$\tilde c_2$  & $-0.392297868(2)$       & $-0.446638947(4)$ \\ \hline
$\tilde c_3$  & $0.15857269(4)$         & $0.0580570(1)$ \\ \hline
$\tilde c_4$  & $0.0728826(7)$          & $-0.260094(9)$ \\ \hline
$\tilde c_5$  & $-0.055460(8)$          & $-0.1441(2)$ \\ \hline
$\tilde c_6$  & $-0.00068(4)$           & $0.0084(6)$ \\ \hline
\end{tabular}
\caption{Fit of the effective central charge of the double $A_{n-1}$ TBA to the functional form~\eqref{eq:IdUV} for $n=3,4$.}
\label{tab:identity_fitresult}
\end{table}

\begin{table}[h]
\centering
\begin{tabular}{|c|c|c|}\hline
 & $n=4$ & $n=5$ \\ \hline\hline
$A$           & $0.4999999981(4)$       & $0.400000000(1)$ \\ \hline
$B$  & $-0.0759952(4)$         & $-0.12157(2)$ \\ \hline
$\tilde c_1$  & $2.5(9)\cdot 10^{-13}$  & $8(24)\cdot 10^{-13}$ \\ \hline
$\tilde c_2$  & $-0.199389474(2)$       & $-0.214836246(4)$ \\ \hline
$\tilde c_3$  & $-3.4(5)\cdot 10^{-7}$  & $3(8)\cdot 10^{-8}$ \\ \hline
$\tilde c_4$  & $0.078515(2)$           & $-0.133661(3)$ \\ \hline
$\tilde c_5$  & $0.000129(8)$           & $0.1311(1)$ \\ \hline
$\tilde c_6$  & $0.02549(1)$            & $0.0493(3)$ \\ \hline
\end{tabular}
\caption{Effective central charge fits to the numerically solved $D_n$ TBA system: parameters of~\eqref{eq:sameS_fitfunc} for $n=4,5$.}
\label{tab:sameS_fitresult}
\end{table}

\section{Parafermionic Liouville reflection amplitudes and TBA}\label{app:Lrefl}

In this appendix we summarise the UV limit fits in the $\hat A_n$ models for various CDD parameter $\alpha$. The volume dependence of the effective central charge was used to approximate $P(r)$ from~\eqref{eq:ceffL}, then the Liouville-phase expansion can be fitted using~\eqref{eq:LiuQuant}.

In particular, we generated the pairs $(P(r),P(r) \log\frac{r}{2\pi})$, where $P(r)$ was extracted from the central charge as $P(r)=\sqrt{\left(\frac{3n}{n+2}-c_{\text{TBA}}(r)\right)/24}$ computed from~\eqref{eq:shiftedCDDTBA} in the region $r\in [10^{-20},1]$. Then we fitted a function with a constant and odd powers up to $5$. The resulting fit parameters can be compared with the expansion of the Liouville reflection factor. The results of such fits for $n=2,3,4,5$ are collected in Tables~\ref{tab:n2},~\ref{tab:n3},~\ref{tab:n4} and~\ref{tab:n5} respectively, for several values of $\alpha$.

\begin{table}[htbp]
\centering
\small
\setlength{\tabcolsep}{3pt}
\begin{adjustbox}{max width=\linewidth}
\begin{tabular}{|lrrrrrr|}
\hline

$\alpha$ & $0$ & $0.1$ & $0.2$ & $0.3$ & $0.4$ & $0.5$ \\ \hline\hline

offset TBA ($-\pi/(4Q)$) & $-0.555360367278$ & $-0.555641644895$ & $-0.556484624712$ & $-0.5578867604989$ & $-0.559843851033$ & $-0.562350102284$ \\
offset theory & $-0.555360367269$ & $-0.555641644888$ & $-0.556484624705$ & $-0.5578867604935$ & $-0.559843851028$ & $-0.562350102279$ \\
offset abs.\ error & $8.3\times10^{-12}$ & $7.1\times10^{-12}$ & $6.8\times10^{-12}$ & $5.4\times10^{-12}$ & $5.6\times10^{-12}$ & $4.4\times10^{-12}$ \\\hline\hline

$\delta_1/(4Q)$ TBA & $1.0506431722$ & $1.0773991720$ & $1.10765991799$ & $1.14140378474$ & $1.17859517852$ & $1.2191851892$ \\
$\delta_1/(4Q)$ theory & $1.0506431714$ & $1.0773991714$ & $1.10765991739$ & $1.14140378427$ & $1.17859517805$ & $1.2191851889$ \\
$\delta_1/(4Q)$ abs.\ error & $7.6\times10^{-10}$ & $6.4\times10^{-10}$ & $6.0\times10^{-10}$ & $4.7\times10^{-10}$ & $4.8\times10^{-10}$ & $3.3\times10^{-10}$ \\\hline\hline

$\delta_3/(4Q)$ TBA & $0.801369$ & $0.7981253$ & $0.7884310$ & $0.7724038$ & $0.7502339$ & $0.7221806$ \\
$\delta_3/(4Q)$ theory & $0.801371$ & $0.7981267$ & $0.7884323$ & $0.7724048$ & $0.7502348$ & $0.7221812$ \\
$\delta_3/(4Q)$ abs.\ error & $1.6\times10^{-6}$ & $1.4\times10^{-6}$ & $1.3\times10^{-6}$ & $1.0\times10^{-6}$ & $9.5\times10^{-7}$ & $5.9\times10^{-7}$ \\\hline\hline

$\delta_5/(4Q)$ TBA & $-0.8273$ & $-0.8174$ & $-0.7877$ & $-0.739$ & $-0.6727$ & $-0.5908$ \\
$\delta_5/(4Q)$ theory & $-0.8295$ & $-0.8194$ & $-0.7895$ & $-0.740$ & $-0.6741$ & $-0.5917$ \\
$\delta_5/(4Q)$ abs.\ error & $2.2\times10^{-3}$ & $2.0\times10^{-3}$ & $1.8\times10^{-3}$ & $1.5\times10^{-3}$ & $1.3\times10^{-3}$ & $8.5\times10^{-4}$ \\
\hline

\end{tabular}
\end{adjustbox}
\caption{Reflection amplitude fits in the $\hat A_n$ model for $n=2$. For each value of $\alpha$ in~\eqref{eq:shiftedCDDTBA} we list the offset $-\pi/4Q$ and the first three expansion coefficients $\delta_1/4Q, \delta_3/4Q, \delta_5/4Q$ of~\eqref{eq:LiuQuant} extracted from the TBA (‘TBA’), predicted from the reflection amplitude~\eqref{eq:ReflA} (‘theory’), and their absolute difference.
}
\label{tab:n2}
\end{table}

\begin{table}[htbp]
\centering
\small
\setlength{\tabcolsep}{3pt}
\begin{adjustbox}{max width=\linewidth}
\begin{tabular}{|lrrrrrr|}
\hline
$\alpha$ & $0$ & $0.1$ & $0.2$ & $0.3$ & $0.4$ & $0.5$ \\
\hline
offset TBA ($-\pi/(4Q)$) & $-0.680174761597$ & $-0.680327891275$ & $-0.680787073648$ & $-0.681551690125$ & $-0.682620714340$ & $-0.6839927190$ \\
offset theory & $-0.680174761587$ & $-0.680327891265$ & $-0.680787073639$ & $-0.681551690116$ & $-0.682620714331$ & $-0.6839927189$ \\
offset abs.\ error & $9.9\times10^{-12}$ & $9.4\times10^{-12}$ & $9.5\times10^{-12}$ & $9.1\times10^{-12}$ & $8.9\times10^{-12}$ & $8.6\times10^{-12}$ \\
\hline\hline
$\delta_1/(4Q)$ TBA & $1.6680289175$ & $1.7022311521$ & $1.738169720$ & $1.7758402281$ & $1.8152353911$ & $1.8563450951$ \\
$\delta_1/(4Q)$ theory & $1.6680289168$ & $1.7022311514$ & $1.738169719$ & $1.7758402275$ & $1.8152353905$ & $1.8563450945$ \\
$\delta_1/(4Q)$ abs.\ error & $7.2\times10^{-10}$ & $6.8\times10^{-10}$ & $6.8\times10^{-10}$ & $6.4\times10^{-10}$ & $6.3\times10^{-10}$ & $5.9\times10^{-10}$ \\
\hline\hline
$\delta_3/(4Q)$ TBA & $0.5342465$ & $0.5332846$ & $0.5304042$ & $0.5256207$ & $0.518959$ & $0.5104565$ \\
$\delta_3/(4Q)$ theory & $0.5342475$ & $0.5332856$ & $0.5304051$ & $0.5256215$ & $0.518960$ & $0.5104573$ \\
$\delta_3/(4Q)$ abs.\ error & $1.0\times10^{-6}$ & $9.4\times10^{-7}$ & $9.4\times10^{-7}$ & $8.7\times10^{-7}$ & $8.4\times10^{-7}$ & $7.7\times10^{-7}$ \\
\hline\hline
$\delta_5/(4Q)$ TBA & $-0.3677$ & $-0.3658$ & $-0.359$ & $-0.35011$ & $-0.3365$ & $-0.319$ \\
$\delta_5/(4Q)$ theory & $-0.3686$ & $-0.3666$ & $-0.360$ & $-0.35092$ & $-0.3373$ & $-0.320$ \\
$\delta_5/(4Q)$ abs.\ error & $9.2\times10^{-4}$ & $8.8\times10^{-4}$ & $8.7\times10^{-4}$ & $8.1\times10^{-4}$ & $7.6\times10^{-4}$ & $6.9\times10^{-4}$ \\
\hline
\end{tabular}
\end{adjustbox}
\caption{Same as Table~\ref{tab:n2} for $n=3$}
\label{tab:n3}
\end{table}

\begin{table}[htbp]
\centering
\small
\setlength{\tabcolsep}{3pt}
\begin{adjustbox}{max width=\linewidth}
\begin{tabular}{|lrrrrrr|}
\hline
$\alpha$ & $0$ & $0.1$ & $0.2$ & $0.3$ & $0.4$ & $0.5$ \\
\hline
offset TBA ($-\pi/(4Q)$) & $-0.78539816341$ & $-0.785497628948$ & $-0.7857959500293$ & $-0.786292900312$ & $-0.7869881035176$ & $-0.787881034855$ \\
offset theory & $-0.78539816339$ & $-0.785497628938$ & $-0.7857959500201$ & $-0.786292900303$ & $-0.7869881035111$ & $-0.787881034844$ \\
offset abs.\ error & $1.3\times10^{-11}$ & $9.5\times10^{-12}$ & $9.2\times10^{-12}$ & $8.7\times10^{-12}$ & $6.5\times10^{-12}$ & $1.1\times10^{-11}$ \\
\hline\hline
$\delta_1/(4Q)$ TBA & $2.1953730581$ & $2.2334163360$ & $2.27254568977$ & $2.312759680$ & $2.3540559150$ & $2.3964310626$ \\
$\delta_1/(4Q)$ theory & $2.1953730573$ & $2.2334163354$ & $2.27254568920$ & $2.312759679$ & $2.3540559147$ & $2.3964310619$ \\
$\delta_1/(4Q)$ abs.\ error & $8.5\times10^{-10}$ & $5.9\times10^{-10}$ & $5.7\times10^{-10}$ & $5.4\times10^{-10}$ & $3.7\times10^{-10}$ & $7.0\times10^{-10}$ \\
\hline\hline
$\delta_3/(4Q)$ TBA & $0.4006847$ & $0.40027911$ & $0.3990627$ & $0.397039$ & $0.3942157$ & $0.39059$ \\
$\delta_3/(4Q)$ theory & $0.4006856$ & $0.40027975$ & $0.3990633$ & $0.397040$ & $0.3942161$ & $0.39060$ \\
$\delta_3/(4Q)$ abs.\ error & $9.0\times10^{-7}$ & $6.5\times10^{-7}$ & $6.0\times10^{-7}$ & $5.9\times10^{-7}$ & $4.1\times10^{-7}$ & $7.1\times10^{-7}$ \\
\hline\hline
$\delta_5/(4Q)$ TBA & $-0.2068$ & $-0.20627$ & $-0.20441$ & $-0.20129$ & $-0.19703$ & $-0.19138$ \\
$\delta_5/(4Q)$ theory & $-0.2073$ & $-0.20675$ & $-0.20486$ & $-0.20174$ & $-0.19739$ & $-0.19185$ \\
$\delta_5/(4Q)$ abs.\ error & $5.8\times10^{-4}$ & $4.8\times10^{-4}$ & $4.5\times10^{-4}$ & $4.5\times10^{-4}$ & $3.6\times10^{-4}$ & $4.7\times10^{-4}$ \\
\hline
\end{tabular}
\end{adjustbox}
\caption{Same as Table~\ref{tab:n2} for $n=4$}
\label{tab:n4}
\end{table}

\begin{table}[htbp]
\centering
\small
\setlength{\tabcolsep}{3pt}
\begin{adjustbox}{max width=\linewidth}
\begin{tabular}{|lrrrrrr|}
\hline
$\alpha$ & $0$ & $0.1$ & $0.2$ & $0.3$ & $0.4$ & $0.5$ \\
\hline
offset TBA ($-\pi/(4Q)$) & $-0.878101841393$ & $-0.878173014764$ & $-0.878386500273$ & $-0.878742194197$ & $-0.879239923947$ & $-0.879879448483$ \\
offset theory & $-0.878101841380$ & $-0.878173014750$ & $-0.878386500257$ & $-0.878742194181$ & $-0.879239923931$ & $-0.879879448467$ \\
offset abs.\ error & $1.3\times10^{-11}$ & $1.4\times10^{-11}$ & $1.5\times10^{-11}$ & $1.5\times10^{-11}$ & $1.6\times10^{-11}$ & $1.6\times10^{-11}$ \\
\hline\hline
$\delta_1/(4Q)$ TBA & $2.6765996013$ & $2.7169839283$ & $2.7581367044$ & $2.8000573160$ & $2.8427447427$ & $2.8861975596$ \\
$\delta_1/(4Q)$ theory & $2.6765996006$ & $2.7169839275$ & $2.7581367035$ & $2.8000573151$ & $2.8427447418$ & $2.8861975587$ \\
$\delta_1/(4Q)$ abs.\ error & $7.3\times10^{-10}$ & $8.1\times10^{-10}$ & $9.0\times10^{-10}$ & $9.0\times10^{-10}$ & $9.6\times10^{-10}$ & $9.7\times10^{-10}$ \\
\hline\hline
$\delta_3/(4Q)$ TBA & $0.3205479$ & $0.32034001$ & $0.3197168$ & $0.318679$ & $0.3172305$ & $0.31537214$ \\
$\delta_3/(4Q)$ theory & $0.3205485$ & $0.32034067$ & $0.3197176$ & $0.318680$ & $0.3172313$ & $0.31537294$ \\
$\delta_3/(4Q)$ abs.\ error & $5.9\times10^{-7}$ & $6.6\times10^{-7}$ & $7.5\times10^{-7}$ & $7.6\times10^{-7}$ & $8.1\times10^{-7}$ & $8.0\times10^{-7}$ \\
\hline\hline
$\delta_5/(4Q)$ TBA & $-0.13240$ & $-0.13211$ & $-0.13131$ & $-0.13002$ & $-0.12822$ & $-0.1259$ \\
$\delta_5/(4Q)$ theory & $-0.13272$ & $-0.13246$ & $-0.13169$ & $-0.13041$ & $-0.12862$ & $-0.1263$ \\
$\delta_5/(4Q)$ abs.\ error & $3.2\times10^{-4}$ & $3.5\times10^{-4}$ & $3.8\times10^{-4}$ & $3.8\times10^{-4}$ & $4.0\times10^{-4}$ & $3.9\times10^{-4}$ \\
\hline
\end{tabular}
\end{adjustbox}
\caption{Same as Table~\ref{tab:n2} for $n=5$}
\label{tab:n5}
\end{table}

\section{IR limit fits}\label{app:C}

In this appendix, we collect our infrared fits. Table~\ref{tab:IRfits} contains a detailed account for $A$ and $D$ models, together with their $\hat A, \hat D$ deformations with $\alpha=0$ up to $n=8$. In Tables~\ref{tab:An1} and~\ref{tab:An2} we present the fits for $\hat A_n$ for $n=4,5$ also with non-vanishing $\alpha$ to demonstrate that the additional CDD factor does not affect the $\Phi_{3,1}$ coupling, but shifts the $T \bar T$ coupling by an amount that depends on the parameter $\alpha$.

\begin{table}[]
    \centering
    \footnotesize
    \begin{tabular}{|c|c|}
    \hline
        TBA & $c(r)-c_{\text{IR}}$ \\ \hline \hline
         $A_4$ & $0.57582517(6)r^{-\frac{8}{5}}-1.340580(2)r^{-2}+0.94877(2)r^{-\frac{12}{5}}$ \\
         $\hat A _4$ &  $0.575821763(2)r^{-\frac{8}{5}}+0.0000014(6)r^{-2}+0.9462076(7)r^{-\frac{12}{5}}$ \\ 
         
         \hline
         $D_4$ & $0.575821953(5)r^{-\frac{8}{5}}-0.0000060(2)r^{-2}-0.946133(2)r^{-\frac{12}{5}}$ \\ 
         $\hat D_4$ & $0.57581764(8)r^{-\frac{8}{5}}+1.340617(3)r^{-2}-0.9493(3)r^{-\frac{12}{5}}$ \\ 
         
         \hline \hline
         $A_5$ & $0.226910975(5)r^{-\frac{4}{3}}+0.206042(6)r^{-2}-0.3267128(8)r^{-2}\log r$ \\
         $\hat A_5$ & $0.226909961(5)r^{-\frac{4}{3}}+1.742580(6)r^{-2}-0.3264410(8)r^{-2}\log r$ \\\hline

         $D_5$ & $0.226909484(6)r^{-\frac{4}{3}}-0.206557(8)r^{-2}+0.326777(1)r^{-2}\log r$ \\
         $\hat D_5$ & $0.22690782(2)r^{-\frac{4}{3}}+1.32874(3)r^{-2}+0.327206(3)r^{-2}\log r$ \\

        \hline \hline
         $A_6$ & $0.1193797(3)r^{-\frac{8}{7}}-0.84014(3)r^{-\frac{12}{7}}+1.6478(2)r^{-2}-1.1188(5)r^{-\frac{16}{7}}$\\
         $\hat A_6$ & $0.1193858(1)r^{-\frac{8}{7}}-0.83827(6)r^{-\frac{12}{7}}+3.2989(4)r^{-2}-1.06434(9)r^{-\frac{16}{7}}$\\
         
         \hline
         $D_6$ & $0.119392457(2)r^{-\frac{8}{7}}+0.8423474(4)r^{-\frac{12}{7}}-0.000317(3)r^{-2}-1.184602(6)r^{-\frac{16}{7}}$\\
         $\hat D_6$ & $0.1193767(4)r^{-\frac{8}{7}}+0.84700(8)r^{-\frac{12}{7}}+1.6249(6)r^{-2}-1.061(1)r^{-\frac{16}{7}}$\\

        \hline \hline
         $A_7$ & $0.07203165(1)r^{-1}-0.350166(2)r^{-\frac{3}{2}}-0.1605(1)r^{-2}+0.27852(2)r^{-2}\log r+0.5655(2)r^{-\frac{5}{2}}$\\
         $\hat A_7$ & $0.07203309(2)r^{-1}-0.350433(3)r^{-\frac{3}{2}}+1.5725(2)r^{-2}+0.28290(3)r^{-2}\log r+0.6894(5)r^{-\frac{5}{2}}$\\

        \hline
             $D_7$ & $0.072032665(3)r^{-1}+0.3501820(5)r^{-\frac{3}{2}}-0.18199(3)r^{-2}+0.281973(5)r^{-2}\log r-0.50748(8)r^{-\frac{5}{2}}$\\
         $\hat D_7$ &  $0.0720389(3)r^{-1}+0.34905(2)r^{-\frac{3}{2}}+1.472(1)r^{-2}+0.2997(2)r^{-2}\log r-0.082(3)r^{-\frac{5}{2}}$\\

        \hline \hline
         $A_8$ & $0.04730375(5)r^{-\frac{8}{9}}-0.194054(4)r^{-\frac{4}{3}}+0.9586(2)r^{\frac{-16}{9}}-1.6780(8)r^{-2}+1.003(1)r^{-\frac{20}{9}}$\\
         $\hat A_8$ & $0.04730525(1)r^{-\frac{8}{9}}-0.194213(1)r^{-\frac{4}{3}}+0.96915(5)r^{\frac{-16}{9}}+0.0915(2)r^{-2}+1.0924(3)r^{-\frac{20}{9}}$\\

         \hline
         $D_8$ & $0.04730506(3)r^{-\frac{8}{9}}+0.194593(2)r^{-\frac{4}{3}}+0.9697(1)r^{\frac{-16}{9}}+0.08496(5)r^{-2}-1.4381(6)r^{-\frac{20}{9}}$\\
         $\hat D_8$ & $0.0473122(1)r^{-\frac{8}{9}}+0.19384(1)r^{-\frac{4}{3}}+1.0174(5)r^{\frac{-16}{9}}+1.669(2)r^{-2}-1.073(3)r^{-\frac{20}{9}}$\\

         \hline

    \end{tabular}
    \caption{Infrared central charge fits with $\alpha=0$ in the $A_n,\hat A _n,D_n$ and the $\hat D_n$ TBA systems, with errors of the fits. Results in the $A_n$ and $D_n$ systems agree with those presented in ~\cite{Feveratietal1996PhLB..374...64F}. Compared to the $T\bar T$ deformed case, there is a clear difference in the coefficient of $r^{-2}$, which is generated by $T\bar T$. The differences can be contrasted to the $r^{-2}$ coefficients in the pure $T\bar T$ perturbation, given in the $\tilde c_1$ row in Table~\ref{tab:identityIRfits}. Fit errors are presented in parentheses. The agreement is excellent for small $n$ and gradually worsens for higher $n$. This is probably due to the truncation of the fit with nearby exponents. Moreover, all the fits were carried in the volume range of $200\leq r \leq 2000$. Changing the fit window might also affect the fits.}
    \label{tab:IRfits}
\end{table}

\begin{table}[h]
    \centering
    \begin{tabular}{|c|c|} \hline
        $\alpha$ & $c(r)-c_{\mathrm{IR}}$ \\ \hline
        0 & $0.57582200(8) r^{-\frac{8}{5}} - 0.000032(8) r^{-2} + 0.9476(2) r^{-\frac{12}{5}}$ \\ \hline
        0.25 & $0.57582198(9) r^{-\frac{8}{5}} + 0.380680(8) r^{-2} + 0.9476(2) r^{-\frac{12}{5}}$ \\ \hline
        0.5 & $0.57582199(9) r^{-\frac{8}{5}} + 0.869522(9) r^{-2} + 0.9476(2) r^{-\frac{12}{5}}$ \\ \hline
        $2\pi$ & $0.5758203(6) r^{-\frac{8}{5}} + 716.4397(7) r^{-2} + 0.927(6) r^{-\frac{12}{5}}$ \\ \hline
    \end{tabular}
    \caption{$\alpha$ dependence of fit parameters for $n=4\,;\, r\in[10^4, 10^7]$}\label{tab:An1}
\end{table}

\begin{table}[h]
    \centering
    \begin{tabular}{|c|c|} \hline
        $\alpha$ & $c(r)-c_{\mathrm{IR}}$ \\ \hline
        0 & $0.22690959(6) r^{-\frac{4}{3}} + 1.7352(2) r^{-2} - 0.32572(2) r^{-2} \log r$ \\ \hline
        0.25 & $0.22690960(6) r^{-\frac{4}{3}} + 2.1723(2) r^{-2} - 0.32572(2) r^{-2} \log r$ \\ \hline
        0.5 & $0.22690960(6) r^{-\frac{4}{3}} + 2.7335(2) r^{-2} - 0.32573(2) r^{-2} \log r$ \\ \hline
        $2\pi$ & $0.22691024(3) r^{-\frac{4}{3}} + 824.1837(7)r^{-2} - 0.32628(6) r^{-2} \log r$ \\ \hline
    \end{tabular}
    \caption{$\alpha$ dependence of fit parameters for $n=5\,;\ r\in[10^4, 10^7]$ }\label{tab:An2}
\end{table}



\newpage

\bibliographystyle{JHEP}
\bibliography{biblio.bib}

\end{document}